\documentclass[preprint,12pt]{elsarticle}

\usepackage{amssymb}
\usepackage{amsmath}
\usepackage{siunitx}
\usepackage{xfrac}  

\usepackage{tikz}
\usetikzlibrary{shapes,positioning,calc,intersections, angles, quotes, snakes}
\tikzset{cross/.style={cross out, draw=black, minimum size=2*(#1-\pgflinewidth), inner sep=0pt, outer sep=0pt},
cross/.default={1pt}}

\usepackage{enumitem}

\usepackage{multicol}
\usepackage{color-edits}
\addauthor[Pascal]{pascal}{red}
\addauthor[Alexander]{alex}{green}
\addauthor[Jean-Marco]{jm}{blue}
\addauthor[Jan]{jan}{orange}

\usepackage[size=tiny]{todonotes}
\presetkeys{todonotes}{fancyline, color=blue!30}{}

\usepackage{subcaption}

\usepackage{xurl} 

\usepackage{tabularx}

\usepackage[htt]{hyphenat}

\usepackage{listings}
\usepackage{mathtools}

\newcommand{\dif}{\mathrm{d}}

\newcounter{bla}

\journal{Computer Physics Communications}

\begin{document}

\begin{frontmatter}



\title{Improvements in charged lepton and photon propagation for the software PROPOSAL}


\author[a]{Jean-Marco Alameddine\corref{author}}
\author[a]{Johannes Albrecht}
\author[a]{Hans Dembinski}
\author[a]{Pascal Gutjahr}
\author[b]{Karl-Heinz Kampert}
\author[a]{Wolfgang Rhode}
\author[a]{Maximilian Sackel}
\author[b]{Alexander Sandrock}
\author[a]{Jan Soedingrekso}

\cortext[author] {Corresponding author.\\\textit{E-mail address:} jean-marco.alameddine@udo.edu}
\address[a]{TU Dortmund University, Department of Physics, Otto-Hahn-Straße 4a, 44227 Dortmund, Germany}
\address[b]{University of Wuppertal, School of Mathematics and Natural Sciences, Gaußstraße 20, 42119 Wuppertal, Germany}

\begin{abstract}
Accurate particle simulations are essential for the next generation of experiments in astroparticle physics.
The Monte Carlo simulation library PROPOSAL is a flexible tool to efficiently propagate high-energy leptons and photons through large volumes of media, for example in the context of underground observatories.
It is written as a \texttt{C++} library, including a Python interface.
In this paper, the most recent updates of PROPOSAL are described, including the addition of electron, positron, and photon propagation, for which new interaction types have been implemented.
This allows the usage of PROPOSAL to simulate electromagnetic particle cascades, for example in the context of air shower simulations.
The precision of the propagation has been improved by including rare interaction processes, new photonuclear parametrizations, deflections in stochastic interactions, and the possibility of propagating in inhomogeneous density distributions.
Additional technical improvements regarding the interpolation routine and the propagation algorithm are described.
\end{abstract}

\begin{keyword}
Monte-Carlo simulation \sep Muon interaction \sep tau propagation \sep air shower \sep electromagnetic interaction \sep Astroparticle physics

\end{keyword}

\end{frontmatter}


\newpage
{\bf NEW VERSION PROGRAM SUMMARY}

\begin{small}
\noindent
{\em Program Title: PROPOSAL}                                          \\
{\em Developer's repository link:} \url{https://github.com/tudo-astroparticlephysics/PROPOSAL} \\
{\em Licensing provisions:} LGPL  \\
{\em Programming language: C++, Python}                                   \\
{\em Journal reference of previous version:} Computer Physics Communications 242 (2019) 132 \\
{\em Does the new version supersede the previous version?:} Yes \\
{\em Reasons for the new version:} Substantial addition of features. Various bugfixes. \\
{\em Summary of revisions:} The library now also treats photons and has the corresponding processes implemented.
  New parametrizations for photonuclear interaction have been implemented.
  The angular deflection in stochastic energy losses has been implemented in addition to the already
    existing multiple scattering implementation, which has been improved to reduce the runtime.
  The implementation of the Landau-Pomeranchuk-Migdal effect has been corrected.
  The propagation algorithm has been improved, including the support of inhomogeneous density distributions. \\
{\em Nature of problem:} Three-dimensional propagation of charged leptons and
    photons through different media.
  Particles lose energy stochastically by ionization, bremsstrahlung, pair production, and
    photonuclear interaction for charged leptons (including annihilation with atomic electrons for positrons) and Compton scattering, pair production,
    photoelectric effect and photohadronic interaction for photons.
  Additionally, they are deflected while propagating through the medium due to both multiple
     elastic Coulomb scattering as well as deflections in individual stochastic interactions.
  Unstable particles eventually decay, producing secondary particles.
  \\
{\em Solution method:} Monte-Carlo simulation.
  The library samples the next interaction point, the type of interaction process, the energy lost in this interaction process, and the energy lost until this point.
  Particles are propagated until they decay, lose all their kinetic energy (for photons: reach a lower energy limit defined by the validity of the underlying cross sections), or until a user-defined termination criterion is reached.
  For each propagation step, the angular deflection and endpoint shift due to multiple scattering is calculated.

  To improve the performance and deal with the divergence of the bremsstrahlung cross section
    for small photon energies, energy losses below a predefined relative or absolute energy
    threshold are treated continuously.
  The computation time is improved by the use of interpolation tables. \\
   \\

\end{small}

\section{Introduction} \label{sec:intro}

PROPOSAL is a \texttt{C++} and Python software library, simulating the stochastic propagation of high-energy charged leptons using Monte Carlo methods \cite{koehne2013proposal}.
It was initially designed to propagate atmospheric or neutrino-induced muons as well as tau leptons through large volumes in the context of underground experiments, replacing the Java-written simulation software MMC \cite{chirkin04MMC}.
Other existing muon simulation tools are MUSIC \cite{Kudryavtsev09}, MUM \cite{Sokalski01MUM}, and the backward simulation software PUMAS \cite{Niess22}.
While each code has its advantages and specialties, the unique feature of PROPOSAL is its flexibility, allowing a detailed customization of the propagation environment, including a choice for the physics parametrizations themselves.
By now, PROPOSAL has been utilized in a wide range of use cases, including neutrino observatories such as IceCube \cite{Abbasi_2022} and KM3NeT \cite{AIELLO2020107477}, radio neutrino simulations \cite{garcia2020rnog}, or dark matter underground experiments \cite{fedynitch2021pico}.


This article describes the improvements made in PROPOSAL since its last update \cite{dunsch2018proposal}, including photon propagation and the improvement of electron and positron propagation in the context of air shower simulations as the main feature.
In Section \ref{sec:muon_interactions}, improvements of the muon propagation, including the implementation of muon deflections in stochastic interactions, are presented.
In Section \ref{sec:eminteractions}, the implementation of high-energy electron, positron, and photon interactions is described.
Section \ref{sec:secondaries} describes the new options for retrieving individual secondary particles from the existing propagation output.
The inclusion of particle propagation in inhomogeneous density distributions is described in Section \ref{sec:inhomogen}.
Further technical improvements, including a new interpolation method, improvements in the propagation routine, and a description of the current installation process, are discussed in Section \ref{sec:technical} before a conclusion is given in Section \ref{sec:summary}.

\section{Improved Muon Interactions} \label{sec:muon_interactions}

\begin{figure}
        \centering
        \includegraphics[width=\textwidth]{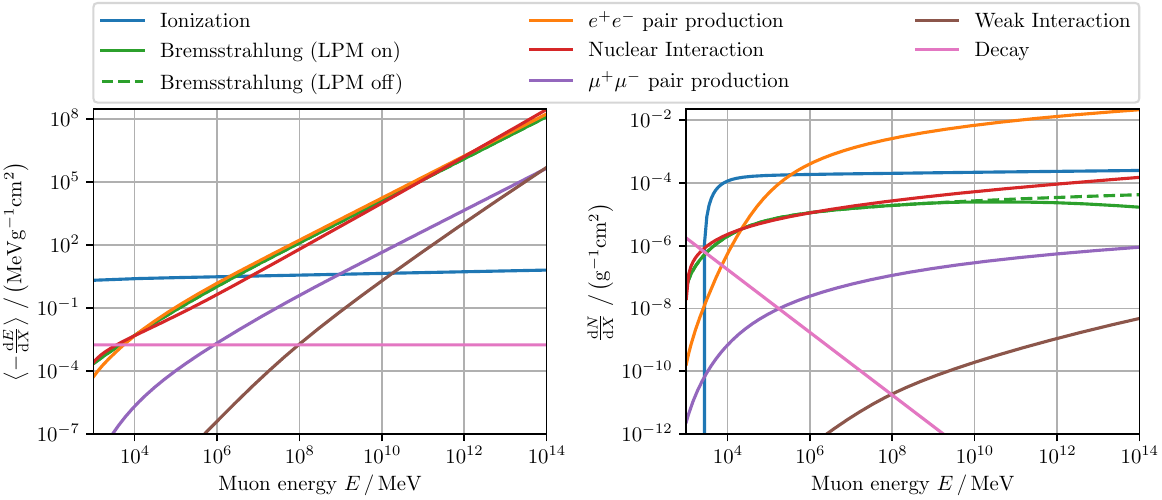}
        \caption{Muon interactions in ice. The left figure shows the average energy loss per grammage, while the right figure shows the number of stochastic interactions. In this example, interactions with an energy loss above \SI{500}{\mega\electronvolt} are treated as stochastic. For all interactions, the default parametrizations (see Table \ref{tab:defaults_muon_tau}) are used.}
        \label{fig:mu_interactions}
\end{figure}

The accuracy of the relevant muon interactions above \si{GeV} energies, i.e., ionization, bremsstrahlung, pair production, and photonuclear interaction, was already on a high level during the initial developments in MMC and has since then continuously been improved.
The theoretical uncertainties of the combined energy loss cross sections are on a percent or sub-percent level, considering even coulomb corrections and one-loop diagrams, see \cite{Sandrock20ICPPA}.
However, due to the large amount of atmospheric muons measured in detectors at the Earth's surface or underground, even rare interactions can occur in significant numbers.
The consideration of rare scenarios and reduction of uncertainties not only for the energy loss but also for the deflection is getting more relevant as detection and reconstruction methods continuously improve.
Therefore, PROPOSAL has improved its muon simulation, including rare interactions, new photonuclear parametrizations, and stochastic deflections, as described in this section.
In addition, an issue in the parametrization of the Landau-Pomeranchuk-Migdal effect has been resolved.
Figure \ref{fig:mu_interactions} shows an overview of all muon energy loss processes currently implemented in PROPOSAL.

\subsection{Rare Interactions: $\mu^+\mu^-$ Pair Production and Weak Interaction}

The production of a $\mu^+\mu^-$ pair only contributes a fraction no more than $\num{\approx e-3}$ to the total energy loss of muons and is, therefore, not a relevant process when describing energy spectra.
However, the produced secondary muons do not deposit their energy near the interaction point but can propagate longer distances through the detector.
In the case of larger transversal momenta of the secondary particles, these multiple muons can be spatially separated from each other, thus producing unique detector signatures.
For small transversal momenta, the muon tracks overlap and are therefore indistinguishable.
This aspect must be considered, e.g., when analyzing the energy loss behavior of muons and muon bundles.
Therefore, muon pair production has been implemented in PROPOSAL, as described in \ref{sec:mupair_xsection}, using the cross section based on \cite{Kelner2000mupair}.
For electrons and taus, this cross section is not applicable.

The weak interaction describes the interaction of a charged lepton, exchanging a W boson with a nucleus while transforming into a neutrino.
This interaction is extremely rare compared to all other processes, as Figure \ref{fig:mu_interactions} indicates.
However, if a muon undergoes a weak interaction inside a detector, its track will disappear and end in a large hadronic cascade, as the track of the outgoing neutrino is not detectable.
Notably, this specific signature might be a background for tau neutrino searches.
Therefore, this process has been implemented in PROPOSAL using the cross sections described in \ref{sec:weak_xsection}, using the cross section based on \cite{CSMS11NuXsection}.
Note that the interaction with the neutral $Z$ boson has not been implemented, as the muon will not disappear, so no significant detector signature is produced.

\subsection{New Photonuclear Parametrizations} \label{sec:new_muon_params}

Photonuclear interactions have the largest theoretical uncertainties regarding interactions of high-energy muons ($\SI{\approx 10}{\%}$).
While this uncertainty is of minor importance for muons around \si{TeV} energies, where photonuclear interactions contribute only \SI{10}{\%} to the overall energy loss, this contribution increases with the muon energy.
At energies above \si{EeV}, most energy is lost through photonuclear interactions, as visible in Figure \ref{fig:mu_interactions}.
Therefore, increasing the accuracy of photonuclear interactions is essential to improve high-energy muon propagation.

\begin{figure}
    \centering
    \includegraphics[width=0.8\textwidth]{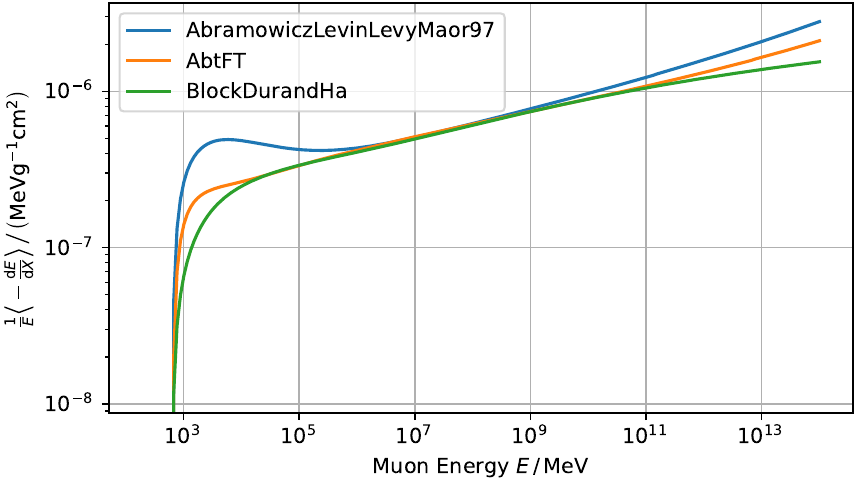}
    \caption{Comparison of the average muon energy loss in ice due to photonuclear interactions. The two new parametrizations in comparison to the current default cross section in PROPOSAL are shown.}
    \label{fig:photonucl_compare}
\end{figure}

However, at these energies, the cross sections can not be validated with experimental measurements and depend on the underlying models.
Therefore, two further parametrizations have been implemented.
The \textit{AbtFT} parametrization \cite{abt2017photonucl} fits the model of the ALLM parametrization, the current default cross section, to more recent measurements.
The \textit{BlockDurandHa} parametrization \cite{block2014photonucl} has the advantage compared to the ALLM models, that it by design obeys the Froissart bound.
Both parametrizations are also applicable for electron or tau propagation.
The cross sections of both parametrizations are described in \ref{sec:newphotonucl_params}, and a visual comparison with the current default cross section of PROPOSAL is shown in Figure \ref{fig:photonucl_compare}.

\subsection{Deflection of Stochastic Interactions} \label{sec:deflection}

\begin{figure}
    \centering
    \includegraphics[width=0.4\textwidth]{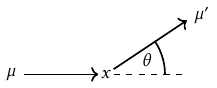}
    \caption{An incoming muon is deflected by an arbitrary interaction $x$. The outgoing muon is deflected by the angle $\theta$ \cite{gutjahr21masters}.}
    \label{fig:defl_single}
\end{figure}

\begin{figure}
    \centering 
    \includegraphics[width=0.85\textwidth]{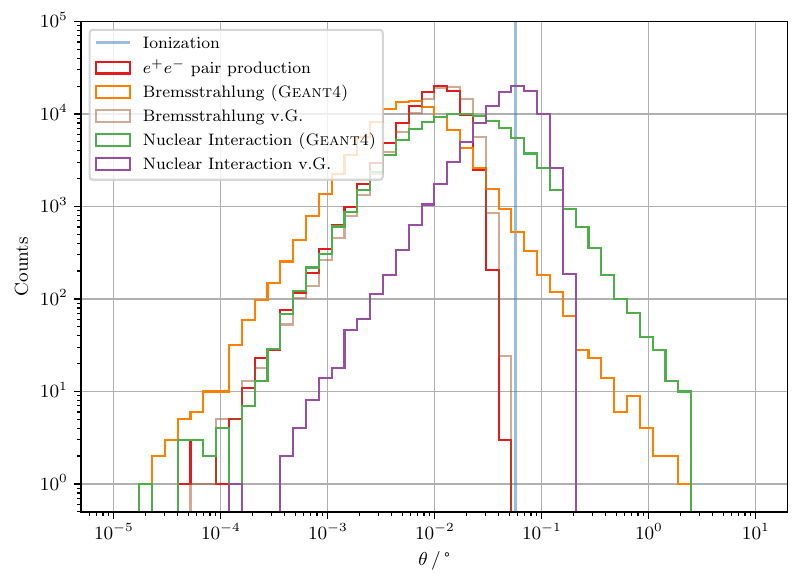}
    \caption{For all interaction types relevant for muons, \num{100000} angles are sampled for each 
    deflection parametrization shown in Figure~\ref{fig:all_deflections_rms}. The muon energy is set to 
    \SI{1}{\tera\electronvolt}, and an energy loss of \SI{50}{\percent} is applied. Due to different sampling 
    methods between Van Ginneken (v.G.) and \textsc{Geant4}, outliers to larger deflections are neglected in 
    Van Ginneken, but the mode of the distribution is shifted to larger angles in Van Ginneken. A single deflection can 
    extend over several orders of magnitude for fixed settings.}
    \label{fig:all_deflections_sampled}
\end{figure}

In addition to multiple scattering, i.e., lateral particle displacement during a continuous step, particles can be deflected in individual stochastic interactions, as illustrated in Figure \ref{fig:defl_single}. This process is called a \emph{stochastic deflection} in the following.
Stochastic deflections of muons for ionization, electron-positron pair production, bremsstrahlung, and photonuclear interaction have been implemented.
For deflections in photonuclear interactions and bremsstrahlung, parametrizations using either the calculation of 
Van Ginneken \cite{vanginneken1986} or as implemented 
in \textsc{Geant4} \cite{GEANT4standard, GEANT4} are provided. The parametrization for electron-positron production is also based on the calculations by 
Van Ginneken, and the deflection angle due to ionization is directly calculated using conservation of four-momentum. 
For each parametrization, it is observed that the lower the energy and the larger the energy loss, the larger the 
deflection. The parametrizations are described in detail in \ref{sec:deflection_params}, and an overview is presented in Table~\ref{tab:deflection_params_muon}.
Figure~\ref{fig:all_deflections_sampled} shows the distribution of individual stochastic muon deflections for the different parametrizations and interaction types.
For each parametrization, except ionization, deflection angles extend over several orders of magnitude and angles larger than \SI{1}{\degree} are possible.


\begin{figure}
    \centering 
    \includegraphics[width=0.85\textwidth]{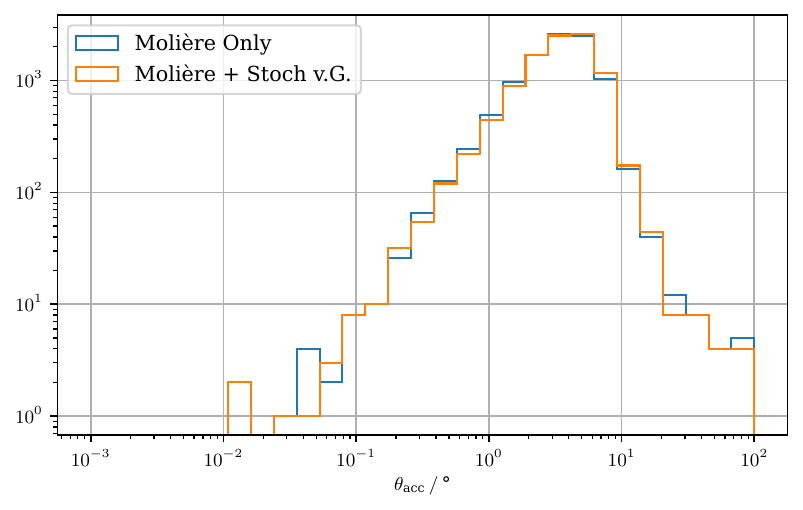}
    \caption{Accumulated deflection $\theta_{\mathrm{acc}}$ between the ingoing and outgoing muon direction. In total, \num{10000} muons are propagated in ice with an initial energy of \SI{1}{\tera\electronvolt} to 
    a final muon energy of \SI{1}{\giga\electronvolt}. A relative energy cut of $v_{\mathrm{cut}} = \num{e-3}$ 
    is used. A comparison between the simulation using only multiple scattering, according to Molière, and a simulation using both multiple scattering and stochastic deflections, according to Van Ginneken (v.G.), is shown.}
    \label{fig:prop_deflections}
\end{figure}

During particle propagation, there is not only one interaction but a multitude of individual interactions, depending on the particle energy and the propagation distance. In each interaction, the particle is deflected by an angle $\theta$. 
All of these single deflections, in addition to the displacement due to multiple scattering, accumulate and lead to a total deflection angle $\theta_{\mathrm{acc}}$, which compares the direction of the muon before and after the propagation.
A histogram of total deflection angles $\theta_{\mathrm{acc}}$, once for particle propagation with and once without stochastic deflections, is presented in Figure ~\ref{fig:prop_deflections}.
Both distributions agree without a significant difference.
However, since large stochastic deflection can occur in individual events, it is necessary to include this effect in simulations, e.g., for experiments reconstructing muon tracks with a high angular resolution.
The impact of stochastic deflections and multiple scattering on the directional reconstruction for neutrino telescopes, like the IceCube Neutrino Observatory, are analyzed in detail in \cite{gutjahr21masters, gutjahr22deflectionpaper}.
In addition, a comparison of the accumulated muon deflection between data and PROPOSAL simulations is shown in \cite{gutjahr22deflectionpaper}.


To be able to estimate the effect of possible uncertainties in deflection parametrizations, an option to arbitrarily scale stochastic deflection or multiple scattering angles is provided by PROPOSAL.
For this purpose, a multiplier $\zeta$, which can be individually set for each interaction type, is applied to the sampled angle by 
\begin{equation}
    \theta_{\mathrm{s}} = \zeta \cdot \theta\,.
\end{equation}
For multiple scattering, the displacement of the particle relative to the initial axis is scaled.

So far, the deflection parametrizations are only available for muons.
Further parametrizations for electrons and positrons need to be implemented in the future.

\subsection{Correction of the Landau-Pomeranchuk-Migdal effect}
\label{sec:lpm}

\begin{figure}
    \centering
    \includegraphics[width=0.8\textwidth]{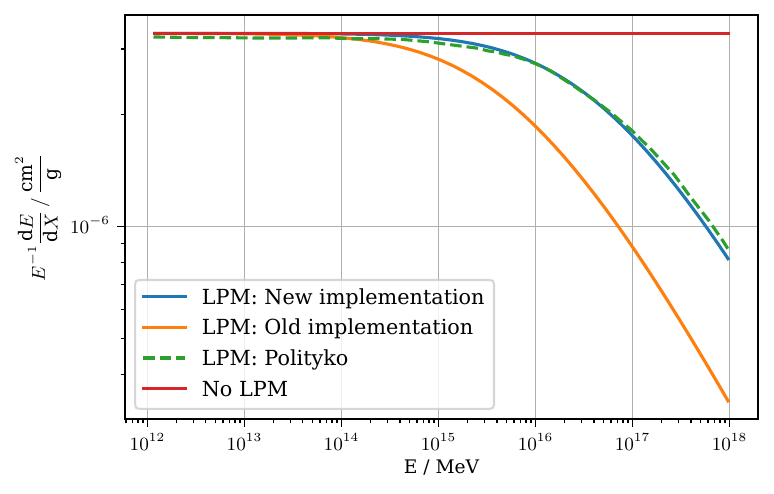}
    \caption{Average energy loss of muons in iron due to bremsstrahlung, with the LPM effect implementation in the old and new PROPOSAL version. As a validation, calculations from \cite{Polityko_2002} are shown. Without the LPM effect, the old implementation, new implementation, and the calculations from \cite{Polityko_2002} are identical.}
    \label{fig:polityko}
\end{figure}

The Landau-Pomeranchuk-Migdal (LPM) effect describes a suppression of small bremsstrahlung losses and pair production processes with a symmetric energy distribution, which is relevant for very-high energies or in dense media.
Within the Migdal functions, an energy scale $E_\text{LPM}$ appears, which can be interpreted as the transition point above which LPM suppression sets in.
Different conventions for this quantity exist, which differ by a factor of 8, leading to different formulae for the Migdal functions.
In earlier versions, different conventions were mixed, such that the LPM suppression of bremsstrahlung started at too low energies.
A careful comparison to literature results on LPM suppression in electron and muon processes \cite{StanevLPM,Polityko_2002} confirms the correctness of the new implementation.
Exemplarily, Figure \ref{fig:polityko} shows the effect of these improvements regarding muon energy losses in iron.

\section{Implementation of high-energy electron, positron, and photon interactions} \label{sec:eminteractions}

Originally, PROPOSAL has been designed for muon and tau propagation in large, homogeneous media.
However, the astroparticle physics community needed a modern and modular interaction module that describes all types of electromagnetic particles.
Since the description of electron, positron, and photon interactions relies on similar approaches from both a methodological and physical point of view, PROPOSAL has been updated to include the necessary parametrizations.
This extension allows the library to be used for the physics description of the electromagnetic and muonic component particle cascades, for example, in the air shower simulation framework CORSIKA~8 \cite{corsika8icrc21, Engel:2018akg}.

\begin{figure}
    \centering
    \begin{subfigure}[t]{0.49\textwidth}
        \centering
        \includegraphics[width=\textwidth]{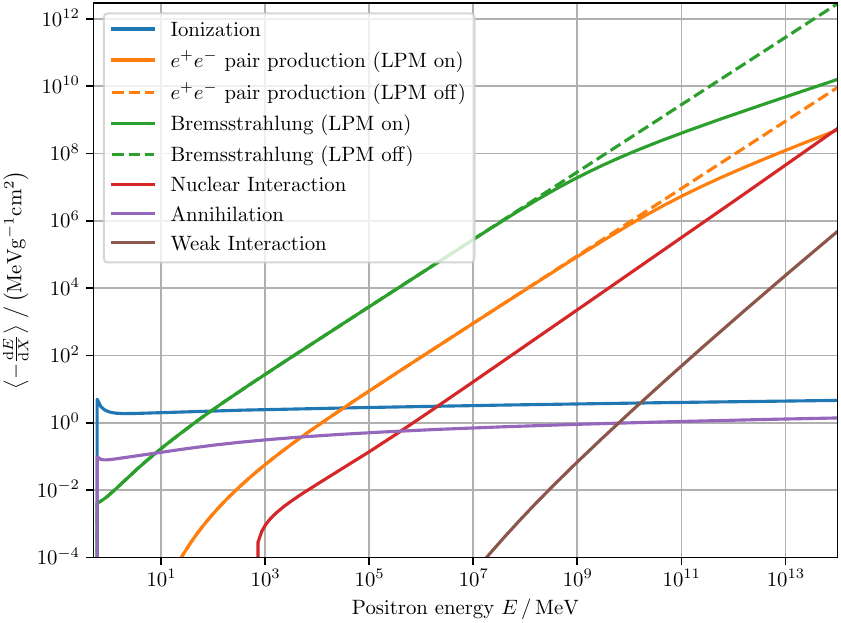}
        \caption{Average positron energy loss per grammage.}
        \label{fig:positron_dedx}
    \end{subfigure}
    \hfill
    \begin{subfigure}[t]{0.49\textwidth}
        \centering
        \includegraphics[width=\textwidth]{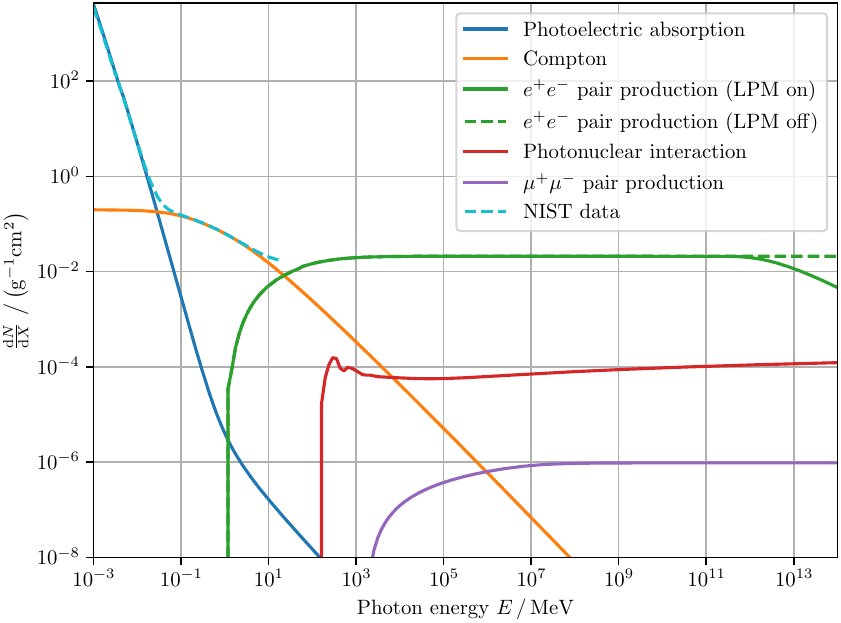}
        \caption{Cross section of photons. As a comparison, the photon cross section according to the NIST Standard Reference Database is shown \cite{nist}.}
        \label{fig:photon_dndx}
    \end{subfigure}
    \caption{Cross sections of positron and photon interactions in air, described by the default parametrizations given in \ref{sec:default_cross}.}
    \label{fig:em_xsections}
\end{figure}

For electrons and positrons, new parametrizations to describe ionization and bremsstrahlung interactions have been implemented.
Furthermore, an adapted parametrization for electron-positron pair production is now available.
As it is relevant for positrons at lower energies and essential for describing the charge excess of electrons over positrons, annihilation has been added as a new interaction type.
For the description of multiple scattering, photonuclear interactions, weak interaction, and the LPM effect on bremsstrahlung, the parametrizations already available in PROPOSAL can be directly applied for electrons and positrons \cite{koehne2013proposal, dunsch2018proposal}.
The details of these parametrizations and their usage in PROPOSAL are described in \ref{sec:ep_params}.
Figure \ref{fig:positron_dedx} shows the average energy loss of positrons in air, using all interaction types available in PROPOSAL.
By default, PROPOSAL uses the parametrizations summarized in \ref{sec:default_cross}.

For the description of photons, the most crucial interaction types are photoelectric absorption (dominant in air at energies below $\approx \SI{30}{\kilo\electronvolt}$), Compton scattering (dominant in air at energies below $\approx \SI{20}{\mega\electronvolt}$), and electron-positron pair production (dominant at higher energies).
Therefore, these interaction types have been implemented in PROPOSAL.
In addition, muon pair production and photonuclear interactions have been implemented as additional processes.
While their contribution to the mean free path of photons is negligible, they are the processes responsible for the production of hadrons and muons in electromagnetic showers, with photonuclear interaction producing most of the lower energetic muons, while muon pair production produces most of the higher energetic muons \cite{HeckMuPair09}.
The photon parametrizations mentioned above and their implementation in PROPOSAL are described in \ref{sec:gamma_params}.
Figure \ref{fig:photon_dndx} shows the cross section of photons in air.

The validity of PROPOSAL as an electromagnetic interaction model has been shown by comparing extensive air showers simulated with CORSIKA~8, using PROPOSAL as a model to describe the electromagnetic and muonic shower component, to air showers simulated with established frameworks such as CORSIKA~7, which uses an adapted version of the code EGS4 \cite{egs4} to describe the electromagnetic shower component \cite{CORSIKA8:2021ilo,icrc2023}.
For relevant shower parameters such as the longitudinal and lateral shower development, an agreement of \SI{10}{\percent} or better, depending on the exact simulation settings, can be observed.
Furthermore, the simulation of radio signals in air showers shows a good agreement, which is highly dependent on a correct description of the electromagnetic component and, therefore, a sensitive test for PROPOSAL \cite{icrc2023_radio}.

\section{Calculation of secondary particles} \label{sec:secondaries}
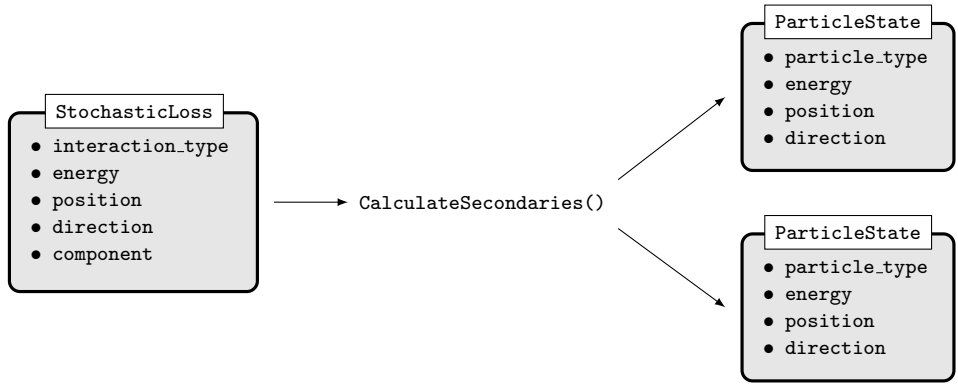
\begin{figure}
\centering
\begin{tikzpicture}[tips=proper]

\tikzstyle{mybox} = [draw=black, fill=gray!20, very thick,
    rectangle, rounded corners, inner sep=4pt, inner ysep=10pt]
\tikzstyle{fancytitle} =[draw=black, fill=white, text=black]

\node [mybox] (stochasticloss){%
    \begin{minipage}{0.22\textwidth}
    	\scriptsize
        \begin{itemize}[wide=4pt,labelsep=4pt]
        	\setlength\itemsep{-0.5em}
        	\item \texttt{interaction\_type}
        	\item \texttt{energy}
        	\item \texttt{position}
        	\item \texttt{direction}
        	\item \texttt{component}
        \end{itemize}
    \end{minipage}
};
\node[fancytitle, right=14pt] at (stochasticloss.north west) {\scriptsize \texttt{StochasticLoss}};

\node[text width=3cm] at (4.5,0) (text) {\scriptsize \texttt{CalculateSecondaries()}};

\node at (9.5,1.4) [mybox] (particlestate1){%
    \begin{minipage}{0.185\textwidth}
    	\scriptsize
        \begin{itemize}[wide=4pt,labelsep=4pt]
        	\setlength\itemsep{-0.5em}
        	\item \texttt{particle\_type}
        	\item \texttt{energy}
        	\item \texttt{position}
        	\item \texttt{direction}
        \end{itemize}
    \end{minipage}
};
\node[fancytitle, right=9pt] at (particlestate1.north west) {\scriptsize \texttt{ParticleState}};

\node at (9.5,-1.4) [mybox] (particlestate2){%
    \begin{minipage}{0.185\textwidth}
    	\scriptsize
        \begin{itemize}[wide=4pt,labelsep=4pt]
        	\setlength\itemsep{-0.5em}
        	\item \texttt{particle\_type}
        	\item \texttt{energy}
        	\item \texttt{position}
        	\item \texttt{direction}
        \end{itemize}
    \end{minipage}
};
\node[fancytitle, right=9pt] at (particlestate2.north west) {\scriptsize \texttt{ParticleState}};

\draw[-latex] ([xshift=0.2cm]stochasticloss.east) edge (text);
\draw[-latex] ([xshift=0.3cm, yshift=+0.3cm]text.east) edge ([xshift=-0.2cm]particlestate1.west);
\draw[-latex] ([xshift=0.3cm, yshift=-0.35cm]text.east) edge ([xshift=-0.2cm]particlestate2.west);

\end{tikzpicture}%
\caption{Visualization of the \texttt{CalculateSecondaries()} method. As an input, the method receives a \texttt{StochasticLoss} object, characterizing the stochastic energy loss of the parent particle. The output is a list of \texttt{ParticleState} objects, where each object describes the properties of a produced secondary particle. Note that the number of created \texttt{ParticleState} objects can be different. In this example, two secondary particles are created.}
\label{fig:calculate_secondaries}
\end{figure}

Initially, PROPOSAL only calculated the energy losses of particles, characterized by their interaction types, energies, and positions.
The directions of these energy losses have not been estimated and have been inherited from their primary particles.
This update includes the functionality to calculate individual secondary particles out of these energy losses, which is necessary to describe particle cascades.
One or several parametrizations to describe this process are provided for each interaction type.
Each parametrization defines a \texttt{CalculateSecondaries()} method, receiving information about the stochastic energy loss, which can, for example, be obtained from the propagator output as explained in Section \ref{sec:propagator_output}, and returns information about the secondary particles.
The technical details of this process are visualized in Figure \ref{fig:calculate_secondaries}.
Note that the calculation of secondary particles is independent of the propagation, i.e., the sampling of energy losses, and therefore optional.
Accordingly, this feature does not affect the performance of applications where only the energy losses are relevant.

Each parametrization needs to solve two related physical tasks: Calculating the energies and the directions of the secondary particles.
The underlying concepts are presented in the following sections.

\subsection{Energy calculation of Secondary Particles}

The energy partition is trivial for interactions with only a single secondary product since the entire energy loss $E \cdot v$ is assigned to the only secondary particle.
Those interactions are bremsstrahlung, ionization, Compton scattering, and photoelectric interactions.
For interactions with two secondary particles that receive energies $E_1$ and $E_2$, where the differential cross section is given in  $x = E_1 / (E_1 + E_2)$, the energy partition can be sampled by solving the integral equation
\begin{align*}
	\frac{1}{\sigma} \int_{x_\text{min}}^{x} \frac{\mathrm{d}\sigma}{\mathrm{d}x^\prime} \, \mathrm{d}x^\prime = \xi
\end{align*}
for $x$, with a random number $\xi \in [0, 1)$ and the total cross section $\sigma$.
This approach is used for annihilation, electron-positron pair production, and muon pair production.
For electron-positron and muon pair production by leptons, the parametrization is described as a double-differential cross section (see Section \ref{sec:electron_positron_pairproduction}).
However, the utilized approach is identical, with the second variable being fixed.

An extra treatment is necessary for the photonuclear interaction of photons and leptons.
The complexity of these interactions, which exceeds the scope of PROPOSAL, requires the usage of hadronic event generators, such as \textsc{Sibyll} \cite{PhysRevD.102.063002} or SOPHIA \cite{2000CoPhC.124..290M}, to sample secondary particles.
Therefore, a hadronic pseudo-particle is returned, whose information can be passed to a hadronic event generator.
This approach is, for example, used in the air shower simulation framework CORSIKA~8 \cite{icrc2023, Engel:2018akg}, where for hadronic energy losses, the energy lost in the interaction as well as the information about the involved target nucleus is passed to either \textsc{Sibyll} or SOPHIA, depending on the chosen transition energy between the low and high-energy hadronic interaction model.
For a more consistent treatment, a sampling of the transferred momentum for cross sections that are differential in $Q^2$ could be implemented in a future update.


\subsection{Directional calculation of Secondary Particles}

For interactions that can be approximated as a two-body interaction with an atomic electron at rest, assuming both energy and momentum conservation, the polar angles of the secondary particles can be calculated deterministically.
In this case, only the energies of the secondary particles need to be known.
This approximation is used for ionization, annihilation, and Compton scattering.
The azimuth angle is sampled uniformly for one secondary particle between $0$ and $2 \pi$, with the other particle receiving the opposite azimuth.

For electron-positron pair production and the photoelectric effect, the assumption that the secondary particles inherit the direction of the parent particle is made.
A more sophisticated description might be added in a future update.

Since bremsstrahlung and electron-positron pair production are the predominant processes for high-energy electromagnetic cascades, their precise treatment is of particular importance.
As a first approximation, the assumption of a polar angle $\theta = m_e / E_0$ can be used, where $E_0$ denotes the energy of the initial particle.
It is possible to use such a simple description since for high energies, the particle production is strongly peaked towards the forward direction, while for lower energies, multiple scattering effects dominate the lateral behavior of the electromagnetic cascade \cite{egs4}.

\begin{figure}
    \centering
    \begin{subfigure}[t]{0.49\textwidth}
        \centering
        \includegraphics[width=\textwidth]{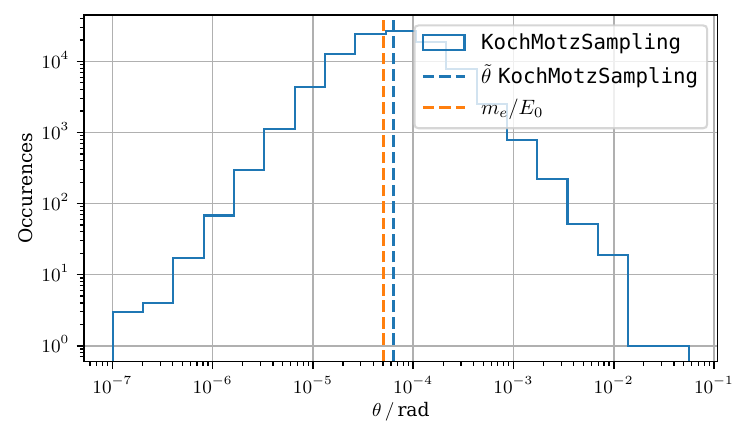}
        \caption{Production angles of bremsstrahlung photons created by electrons or positrons. The plot shows a comparison of the $\theta = m_e / E_0$ approximation with the parametrization \texttt{KochMotzSampling}, described in \ref{sec:secondaries_brems}.}
        \label{fig:brems_angle}
    \end{subfigure}
    \hfill
    \begin{subfigure}[t]{0.49\textwidth}
        \centering
        \includegraphics[width=\textwidth]{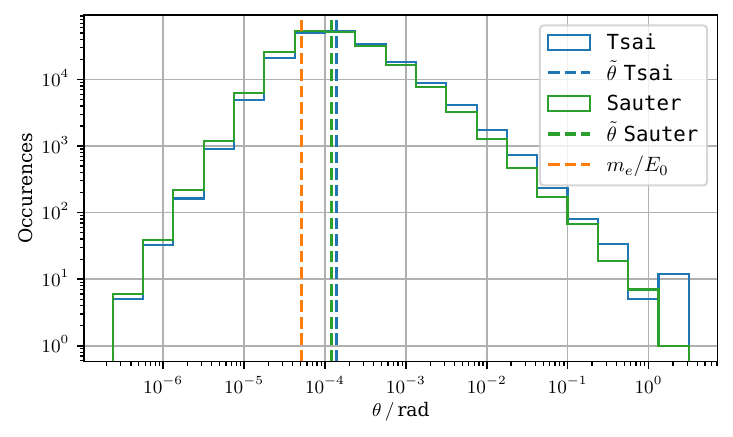}
        \caption{Production angles of electron and positrons created by photons. Compared are the $\theta = m_e / E_0$ approximation , the \texttt{Tsai} parametrization according to \eqref{eqn:angular_tsai}, and the \texttt{Sauter} sampling procedure according to \eqref{eqn:sauter_sampling}.}
        \label{fig:photopair_angle}
    \end{subfigure}
    \caption{Production angles of secondary particles in air, with an initial particle energy of $E = \SI{e4}{\mega\electronvolt}$. The variable $\tilde{\theta}$ indicates the median of the corresponding distribution.}
    \label{fig:production_angles}
\end{figure}

As an alternative, a rejection sampling method for the bremsstrahlung angle distribution, based on the double differential cross section by Koch and Motz \cite{RevModPhys.31.920}, has been implemented.
The details of this method are described in \ref{sec:secondaries_brems}.
Figure \ref{fig:brems_angle} compares both approaches to sample the production angle of bremsstrahlung photons.
The direction of the outgoing electron or positron is calculated assuming momentum conservation and neglecting the transferred momentum to the nucleus in the bremsstrahlung interaction.

For electron-positron pair production, two additional alternatives to the $\theta = m_e / E_0$ approximation have been implemented.
These methods are based on differential cross sections describing the angular distribution of the produced leptons.
The first method, \texttt{Tsai}, is based on the parametrization described in \cite{tsai1974}, while the second method, \texttt{Sauter}, is based on a parametrization given in \cite{RevModPhys.41.581}.
Both methods are described in \ref{sec:secondaries_epair}.
Figure \ref{fig:photopair_angle} compares all three sampling approaches.
\section{Propagation in Media with Inhomogeneous Densities} \label{sec:inhomogen}

This update introduces the possibility of propagating particles through inhomogeneous density distributions, allowing for a more detailed description of the environment, for example, the overburden of underground experiments or the atmosphere of the Earth.
To make this possible, cross sections and the corresponding interaction integrals are now internally calculated in terms of grammage instead of distances, making them independent of the actual density distribution.
Assuming the density is a linear factor in the cross section, the density correction can be separated from the average energy loss, i.e.
\begin{equation}
    \label{eqn:density_linear_factor}
    f(E) \coloneqq \frac{\dif E}{\dif X} = \frac{\dif E}{\dif x} \frac{1}{\rho(x)}.
\end{equation}
This integral can be solved by separation of variables
\begin{equation}
    \int \frac{1}{f(E)} \dif E = \int \rho(x) \dif x,
\end{equation}
describing the relation between energies and distances given a density distribution $\rho(x)$.
Four different options to describe $\rho(x)$ have been implemented:
A homogeneous distribution, an exponential distribution, a polynomial distribution, and the description of the distribution via splines.
The density inside a geometry can be varied along a Cartesian or a radial axis.
For a given origin $\vec{f}_{p_0}$, the depth $d(\vec{x})$ of a radial axis is calculated by
\begin{equation}
    d(\vec{x}) = |\vec{x} -\vec{f}_{p_0}| .
\end{equation}
For a Cartesian axis, an additional direction $\vec{f}_\mathrm{axis}$ needs to be defined, and $d(\vec{x})$ is calculated by
\begin{equation}
    d(\vec{x}) = \vec{f}_\mathrm{axis} \cdot (\vec{x} -\vec{f}_{p_0}) .
\end{equation}
The exponential distribution is then defined by the scaling parameter $\sigma$ and the shifting parameter $d_0$, resulting in the density profile
\begin{equation}
    \rho(\vec{x}) = \rho_0 \cdot \exp{\left(\frac{d(\vec{x}) - d_0}{\sigma}\right)} .
\end{equation}
The polynomial distribution is described by the coefficients $a_k$, resulting in the density profile
\begin{equation}
    \rho(x) = \rho_0 \cdot \sum_{k=0}^{n} a_k d(x)^k .
\end{equation}
Either a linear or a cubic spline interpolation can be used to describe the spline density profile.

Treating the density correction as a linear factor as assumed in \eqref{eqn:density_linear_factor} is not valid when the LPM effect (see Section \ref{sec:lpm}) is taken into account, which contains a non-linear density dependency.
However, creating density-independent interpolation tables of the cross sections and propagation integrals requires this assumption.
Therefore, the LPM effect can only be considered using a reference point with a fixed density, assuming no significant density changes.
Since the density effects are mainly minor corrections or have significant effects only at higher energies, this approximation is still valid for most environments.
However, this issue needs to be addressed in a future update, especially in the context of electron, positron, and photon propagation in the atmosphere.


\section{Technical Improvements} \label{sec:technical}

\subsection{Installation process}

The current dependencies of PROPOSAL are the libraries \texttt{cubic\_interpolation} \cite{cubic_interpolation}, \texttt{spdlog} \cite{spdlog}, \texttt{nlohmann\_json} \cite{Lohmann_JSON_for_Modern_2022}, and \texttt{pybind11} \cite{pybind11}, where the latter is only necessary if python bindings are supposed to be created.
The installation of PROPOSAL is based on CMake.
Instead of manually providing all dependencies, they can be provided by the \texttt{C++} software package manager conan \cite{conan}.
Installation of PROPOSAL as a \texttt{C++} library using conan works via the commands:
\begin{verbatim}
 $ pip install conan
 $ git clone \
 $ https://github.com/tudo-astroparticlephysics/PROPOSAL.git
 $ cd PROPOSAL
 $ conan install . --build=missing
 $ cd build
 $ cmake .. -DCMAKE_TOOLCHAIN_FILE=conan_toolchain.cmake
 $ cmake --build . -j
 $ cmake --install .
\end{verbatim}
The step \verb|conan install| provides the dependencies by either fetching their binaries or building them, which are then passed to CMake, which is used to build and install PROPOSAL in the usual way.

For the usage of PROPOSAL as a Python package, versions later than \texttt{v6.1.0} can be installed via the Python software repository PyPi \cite{pypi}, with the command:
\begin{verbatim}
 $ pip install proposal
\end{verbatim}

In addition to Linux and macOS, PROPOSAL is now also developed and tested for Windows.
A detailed description of the installation process is provided in \url{https://github.com/tudo-astroparticlephysics/PROPOSAL/blob/master/INSTALL.md}.

\subsection{Modularization}

As a central feature, PROPOSAL provides a complete three-dimensional Monte Carlo simulation of individual particles.
Given an input, where the initial particle state and the propagation environment are defined, PROPOSAL returns information about the energy losses and particle states, where the output is explained in detail in Section \ref{sec:propagator_output}.
This functionality is provided by the \texttt{Propagator} class.

Several calculations regarding the continuous and stochastic interaction have to be performed to execute a propagation step.
For this update of PROPOSAL, these tasks have been moved from inside the \texttt{Propagator} class to individual modules, which are used by the \texttt{Propagator} but can also be used as standalone classes.
This restructuring allows PROPOSAL to be used also as a modular library. 
Six different modules are available: The \emph{Interaction}, \emph{Decay}, \emph{ContinuousRandomization}, \emph{Displacement}, \emph{Time}, and \emph{Scattering} module. \cite{Alameddine:2020zyd}

The \emph{Interaction} module provides the method to calculate the energy $E_\text{f}$ of the next stochastic interaction, which is defined as an interaction with a relative energy loss above a selected energy cut $v_\text{cut}$.
To calculate $E_\text{f}$, the integral equation
\begin{equation}
	\label{eqn:energy_integral}
	\int_{E_\text{i}}^{E_\text{f}} \frac{\sigma(E)}{-f(E)} \, \mathrm{d}E = - \log{\xi}
\end{equation}
is solved, where $\sigma(E)$ is the total stochastic cross section, $f(E) = \mathrm{d}E / \mathrm{d}X$ the average energy loss per grammage, and $\xi \in [0, 1)$ a random number \cite{koehne2013proposal}.
Furthermore, the type and size of a stochastic energy loss can be sampled from the differential cross section.
Lastly, the individual integrated cross sections of the different processes and their sum can be accessed.
With the \emph{Decay} module, the final energy for which the particle will decay is sampled, given the particle's lifetime and current energy.
The \emph{ContinuousRandomization} module provides a Gaussian smearing of the continuous energy loss, where the mean of the smearing is calculated according to
\begin{equation}
	\left< \Delta (\Delta E)^2 \right> = \int_{E_\text{i}}^{E_\text{f}} \frac{E^2}{-f(E)} \left< \frac{\mathrm{d}^2E}{\mathrm{d}X^2} \right>.
\end{equation}
This method can resolve simulation artifacts, which can occur when using large energy cuts \cite{koehne2013proposal}.
The time elapsed during a propagation step can be calculated with the \emph{Time} module using 
\begin{equation}
	t_\text{f} = t_\text{i} + \int_{x_\text{i}}^{x_\text{f}} \frac{\mathrm{d}x}{v(x)} = t_i - \int_{E_i}^{E_f} \frac{\mathrm{d}E}{f(E)v(E)},
\end{equation}
while the grammage covered during a propagation step with a given initial and final energy is calculated by
\begin{equation}
	\label{eqn:tracking_integral}
	X_\text{f} = X_\text{i} - \int_{E_\text{i}}^{E_\text{f}} \frac{\mathrm{d}E}{f(E)},
\end{equation}
which is provided by the \emph{Displacement} module. 
Additionally, this module calculates the final energy for a given initial energy and grammage, which is the inversion of the integral in \eqref{eqn:tracking_integral}.
Lastly, the \emph{Scattering} module is responsible for describing the lateral particle development.
For the displacement during a continuous propagation step, multiple scattering, using the methods described in Section \ref{sec:scattering_runtime}, can be applied \cite{dunsch2018proposal}. 
The particle deflection during a stochastic interaction is also calculated here, according to the methods described in Section \ref{sec:deflection}. 

One application example of this modular structure is the particle cascade framework CORSIKA~8, which uses the modules provided by PROPOSAL to calculate the electromagnetic and muonic shower component of extensive air showers \cite{corsika8icrc21, Engel:2018akg, CORSIKA8:2021ilo}.
Here, CORSIKA~8 provides the overall algorithm for the Cascade simulation, resorting to physics information from PROPOSAL, for example the mean free path length or the sampling of stochastic interactions (provided by the \emph{Interaction} module), conversions of covered grammages to particle energies and vice versa (provided by the \emph{Displacement} module), or the description of multiple scattering (provided by the \emph{Scattering} module).

\subsection{Improved interpolation routine}

\begin{figure}
    \centering
    \includegraphics[width=0.8\textwidth]{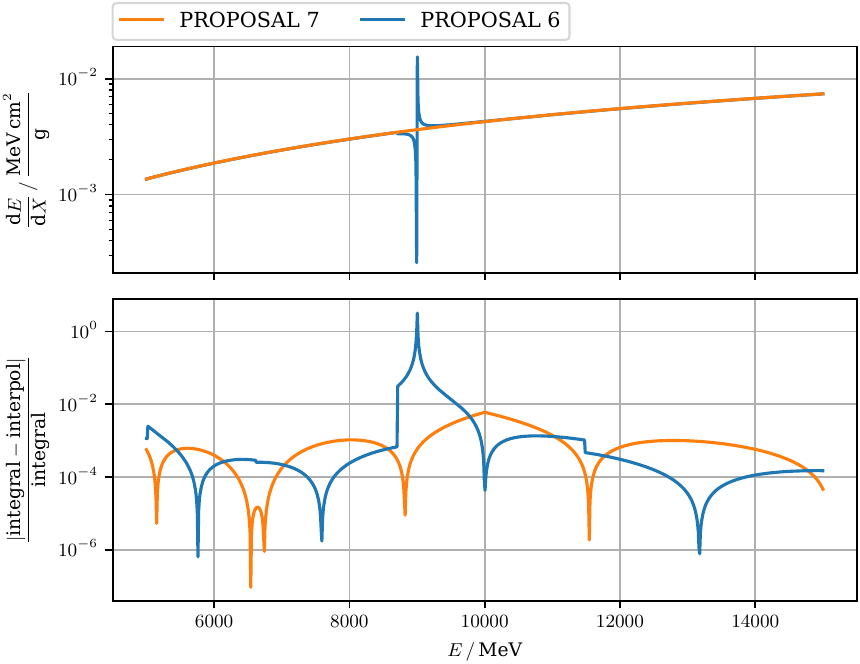}
    \caption{Comparison of the interpolation routine used in PROPOSAL~6 and PROPOSAL~7. Exemplarily, the average pair production energy loss of a muon in ice, using an energy cut of $E_\text{cut}=\SI{500}{\mega\electronvolt}$, $v_\text{cut}=\num{0.05}$, is shown. The upper plot shows the absolute value of the interpolation, the lower plot the ratio between integration and interpolation for PROPOSAL~6 and PROPOSAL~7. Note that for a fair comparison, the number of interpolation nodes has been set to \num{100} in both PROPOSAL versions. Per default, PROPOSAL~7 uses a higher number of interpolation notes to achieve a better precision.}
    \label{fig:interpolation_comparison}
\end{figure}

Several numerically-expensive evaluations of one or two-dimensional integrals are performed during the propagation process.
To speed up this process, PROPOSAL uses interpolation tables to store the results of these integrations.
Previous versions of PROPOSAL used a self-implemented polynomial and rational interpolation routine \cite{koehne2013proposal}.
However, this implementation could show an unstable behavior for functions with non-differentiable parts caused by features present within physics parametrizations or artifacts introduced by the energy cuts.
The upper part of Figure \ref{fig:interpolation_comparison} shows an example of this behavior.

This update of PROPOSAL uses the \texttt{cubic\_interpolation} \cite{cubic_interpolation} library for interpolations.
It provides an implementation of one-dimensional cubic interpolation and two-dimensional bicubic interpolation, based on the Eigen library \cite{eigenweb} and the \texttt{cardinal\_cubic\_b\_spline} class from the Boost.Math library \cite{10.5555/2049814}.
Furthermore, linear and exponential axis transformations and methods to solve integral equations are available.

Figure \ref{fig:interpolation_comparison} compares the performance of the interpolation routines of PROPOSAL~6 and PROPOSAL~7 for a function with a non-differentiable behavior.
Note that while the interpolation routine of PROPOSAL~6 shows an unphysical artifact, the interpolation in PROPOSAL~7 remains at a stable accuracy.

\subsection{Runtime improvements for multiple scattering}
\label{sec:scattering_runtime}

\begin{figure}
    \centering
    \includegraphics[width=\textwidth]{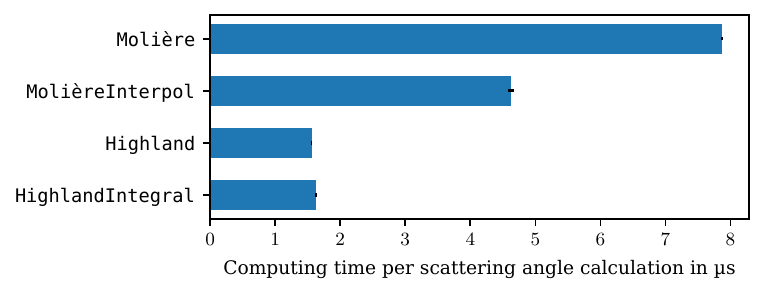}
    \caption{Comparison of the average runtime to calculate a scattering angle according to the different parametrizations implemented in PROPOSAL. A statistic of \num{e7} samples is used.}
    \label{fig:scattering_runtime}
\end{figure}

To describe multiple scattering, different parametrizations to sample the deflection angles are available.
The most accurate description is given by the \texttt{Molière} parametrization, while the \texttt{Highland} and \texttt{HighlandIntegral} parametrizations are a Gaussian approximation of \texttt{Molière} \cite{dunsch2018proposal}.
While they provide an incomplete description of outliers, i.e., large scattering angles, the parametrizations of \texttt{Highland} and \texttt{HighlandIntegral} are much quicker to evaluate, as can be seen in Figure \ref{fig:scattering_runtime}.
However, for some applications, the more accurate description of \texttt{Molière} is necessary, for example, to describe the scattering of electrons and positrons at low energies.
To improve the computing time, the evaluation of two series expansions, which were previously calculated using Horner's method, have been replaced by two interpolation tables.
This method called \texttt{MoliereInterpol} improves the overall evaluation of the scattering angles by a factor of $\approx 2$, as shown in Figure \ref{fig:scattering_runtime}, while providing the same results. 

\subsection{Restructuring of the Propagation Algorithm}
\label{sec:termination_conditions}

With this update of PROPOSAL, the propagation algorithm has been restructured, introducing a more intuitive and consistent structure, also considering the inclusion of non-decaying particles.
A propagation step now consists of two parts: Firstly, the energy of the next interaction point is sampled, and afterward, the distance of the actual propagation step is calculated.
The energy of the next interaction point is defined by
\begin{align} \label{eq:prop_energy_choose}
    E_\text{f} = \max(E_{\text{interaction}}, E_{\text{decay}}, E_{\text{min}}),
\end{align}
where $E_{\text{interaction}}$ is the energy where the next stochastic interaction would occur, sampled according to \eqref{eqn:energy_integral}, $E_{\text{decay}}$ is the sampled energy where the particle would decay, and $E_{\text{min}}$ is the energy threshold where the particle propagation will stop.
This threshold $E_{\text{min}}$ is a new feature of PROPOSAL and can optionally be passed when propagating a particle.

Afterward, the distance $x_\text{f}$ of the actual propagation step is calculated according to
\begin{align}
    x_\text{f} = \min(x_{\text{interaction}}, x_{\text{max}}, x_{\text{border}}),
\end{align}
where $x_{\text{interaction}}$ is calculated by solving \eqref{eqn:tracking_integral} with $E_f$ from \eqref{eq:prop_energy_choose}, $x_{\text{border}}$ is the distance to the next geometry border, and $x_{\text{max}}$ is maximal propagation distance defined by the user.
For $x_\text{f} = x_{\text{interaction}}$, the final energy of the propagation step $E_\text{f}$ is already known, while for the other cases, it needs to be recalculated using \eqref{eqn:tracking_integral}. 


\subsection{Revised output of Propagator class}
\label{sec:propagator_output}


This update of PROPOSAL features an improved interface to the output of a particle propagation process.
During propagation, the \texttt{Propagator} class stores the information about the particle state (i.e., energy, position, direction, time, propagated distance, and most recent interaction type) before and after every stochastic interaction, as well as at sector transitions (which are described in Section \ref{sec:border_transition}).
All particle states are collected in a standard vector. 
In addition to directly accessing this vector, PROPOSAL now provides query methods to intuitively extract specific information about the propagation output.
This includes methods to return lists of all stochastic or continuous energy losses, either in general, for a specific interaction type, or inside a given geometry.
Table \ref{tab:secondaries_method} lists all query methods provided by the \texttt{Secondaries} class.
\begin{table}
	\footnotesize
	\centering
	\caption{Methods provided by the \texttt{Secondaries} class, providing access to the output information of a propagation process.}
	\label{tab:secondaries_method}
	\begin{tabularx}{\textwidth}{l X}
	\hline
	Method & Method description \\
	\hline
	\texttt{GetStochasticLosses()} & Returns list of all stochastic losses during propagation. \\
	\texttt{GetStochasticLosses(geometry)} & Returns list of all stochastic losses inside a geometry. \\
	\texttt{GetStochasticLosses(type)} & Returns list of all stochastic losses of a given interaction type. \\
	\texttt{GetContinuousLosses()} & Returns list of all continuous losses during propagation. \\
	\texttt{GetContinuousLosses(geometry)} & Returns list of all continuous losses inside a geometry. \\
	\texttt{GetInitialState()} & Particle state at start of propagation. \\
	\texttt{GetFinalState()} & Particle state at the end of propagation. \\
	\texttt{GetStateForEnergy(energy)} & Particle state for a given energy. \\
	\texttt{GetStateForDistance(distance)} & Particle state for a given propagation distance. \\
	\texttt{GetEntryPoint(geometry)} & Return particle state when entering geometry, if applicable. \\
	\texttt{GetExitPoint(geometry)} & Return particle state when leaving geometry, if applicable. \\
	\texttt{GetClosestApproachPoint(geometry)} & Return particle state when the smallest distance to the center of geometry has been reached. \\
	\texttt{HitGeometry(geometry)} & Returns true if a geometry has been hit. \\
	\texttt{GetELost(geometry)} & Returns sum of energy that has been deposited in geometry. \\
	\texttt{GetDecayProducts()} & Returns list of decay products, if particle has decayed at the end of propagation. \\
	\hline
	\end{tabularx}
\end{table}
%

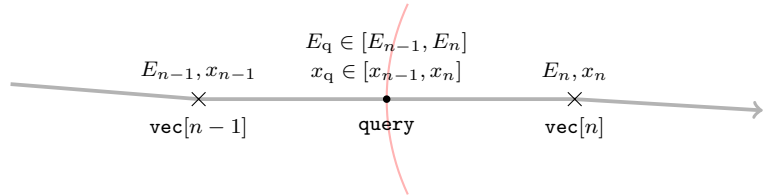
\begin{figure}
	\centering
\begin{tikzpicture}
	\coordinate (start) at (-1, 0.2);
	\coordinate (end) at (9, -0.15);
	\coordinate (n1) at (1.5, 0);
	\coordinate (n2) at (6.5, 0);
	\coordinate (query) at (4, 0);

	\draw [line width=0.5mm, red!30,thick,domain=-25:25] plot ({-3*cos(\x)+7}, {3*sin(\x)});

	\draw [line width=0.5mm, gray!60] (start) -- (n1);
	\draw [line width=0.5mm, gray!60] (n1) -- (n2);
	\draw [line width=0.5mm, gray!60, ->] (n2) -- (end);
	\draw (n1) node[cross=3pt, label=above:{\scriptsize $E_{n-1}, x_{n-1}$}, label=below:{\scriptsize $\texttt{vec}[n-1]$}] {};
	\draw (n2) node[cross=3pt, label=above:{\scriptsize $E_{n}, x_{n}$}, label=below:{\scriptsize $\texttt{vec}[n]$}] {};

	\fill (query) circle (0.05) node[above=0.07] {\scriptsize \shortstack{$E_\text{q} \in [E_{n-1},E_{n}]$\\$x_\text{q} \in [x_{n-1}, x_{n}]$}} node[below=0.15] {\scriptsize \texttt{query}};

\end{tikzpicture}
\caption{Visualization of the concept to calculate particle states at arbitrary query points, given the stored particle states \texttt{vec} with the corresponding particle energies $E_n$ and propagated distances $x_n$. Note that the particle track is assumed to be the direct connection of the stored particle positions. In this example, the query position is defined by the intersection of the particle track with a geometry, indicated by the red circular arc.}
\label{fig:repropagation}
\end{figure}

In addition, the particle state at the exit, entry, or closest approach point of an arbitrary geometry can be obtained.
The underlying algorithm that is used by PROPOSAL is described as follows:
As illustrated in Figure \ref{fig:repropagation}, the algorithm calculates the intersection of the geometry with the particle track, which is given by the direct connection of the positions of the stored particle states.
Then, the closest particle state before the intersection point $\texttt{vec}[n-1]$, with known energy $E_{n-1}$, is determined.
In addition, the propagated distance of this state, $x_{n-1}$, and the propagated distance of the intersection, $x_\text{q}$, are known, which is sufficient to calculate the particle state at the intersection point.
For example, the energy at the intersection point $E_q$ can be calculated by solving the integral equation \eqref{eqn:tracking_integral} for $E_\text{f} = E_\text{q}$, with the already known quantities $x_\text{f} = x_\text{q}$, $x_\text{i} = x_{n-1}$, and $E_\text{i} = E_{n-1}$.
For the conversions of $x$ to $X$, i.e., from distances to grammage and vice versa, the underlying density profile needs to be considered (see Section \ref{sec:inhomogen}).
The same approach can be used to request the particle state for an arbitrary propagation distance or particle energy, using \eqref{eqn:tracking_integral}.

Note that the calculation of the individual decay products is only performed if explicitly requested by the user using the \texttt{GetDecayProducts()} method, similar to the calculation of the secondary particles of an interaction described in Section \ref{sec:secondaries}.
Otherwise, only the decay point is calculated, saving computational resources.
This approach also allows users to sample decay products with other, more sophisticated algorithms, for example, the framework \texttt{TAUOLA} \cite{Chrzaszcz:2016fte}.

\subsection{Treatment of border transitions in the Propagator} \label{sec:border_transition}

\begin{figure}
	\centering
\begin{tikzpicture}
	\draw [thick, double, name path=transition] (-0.5, -0.2) -- node[above, xshift=2.5cm, yshift=0.2cm] {medium A} node[below, xshift=2.5cm, yshift=-0.2cm] {medium B} (7, -0.2);

	\coordinate (A) at (0, 0.5);
	\coordinate (B) at ($(A)+(-20:4)$);
	\coordinate (C) at ($(A)+(0:4)$);

	\fill (A) circle (0.05) node[above] {$\vec{x_\text{i}}$};
	\draw [->] (0, 0.5) -- node[above] {$\vec{d_\text{i}}$} (C);

	\path [name path=line1] (A) -- (B);
	\draw [name intersections={of=transition and line1,by={Int}}] (A) -- (Int);
	\draw [->, red, name intersections={of=transition and line1,by={Int}}] (Int) -- (B);

	\pic [draw, <-, "$\theta$", angle eccentricity=1.5, angle radius=0.9cm] {angle = B--A--C};
\end{tikzpicture}
\caption{Visualization of the border transition problem. A particle at position $\vec{x_\text{i}}$ within medium A is transferred along an initial direction $\vec{d_\text{i}}$. Due to multiple scattering, the initial direction is changed by an angle $\theta$. This can cause the particle to enter a medium B. If this is not correctly taken into account, the particle will traverse a distance through medium B, indicated by the red color, while the algorithm assumes that the particle is still in medium A.}
\label{fig:border_problem}
\end{figure}
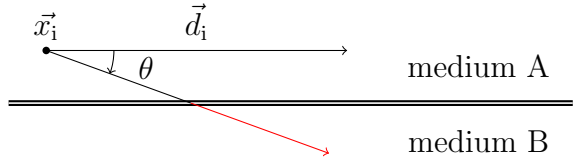

By design, particles simulated with PROPOSAL are allowed to perform large propagation steps.
Since calculations provide different results if the underlying medium is different, the borders between sectors pose an upper limit for propagation steps.
In previous versions of PROPOSAL, during a propagation step, the particle translation due to multiple scattering has been applied after the energy and distance of the step have been calculated.
For particles traveling close to a sector transition, this could lead to problems, as illustrated in Figure \ref{fig:border_problem}.
Due to multiple scattering, particles could enter sectors with other medium properties without this being taken into account by the propagation algorithm.
Especially for particles traveling in parallel to a sector border and for medium transitions with drastic mass density changes, this could lead to under- or overestimation of energy losses and propagation distances.

A naive solution to this problem would be to re-evaluate the distance to the next border transition after calculating the multiple scattering translation and shorten the propagation step if a border transition is violated.
However, since the calculation of the multiple scattering translation is correlated with the distance of a propagation step, this approach would cause an overestimation of multiple scattering.

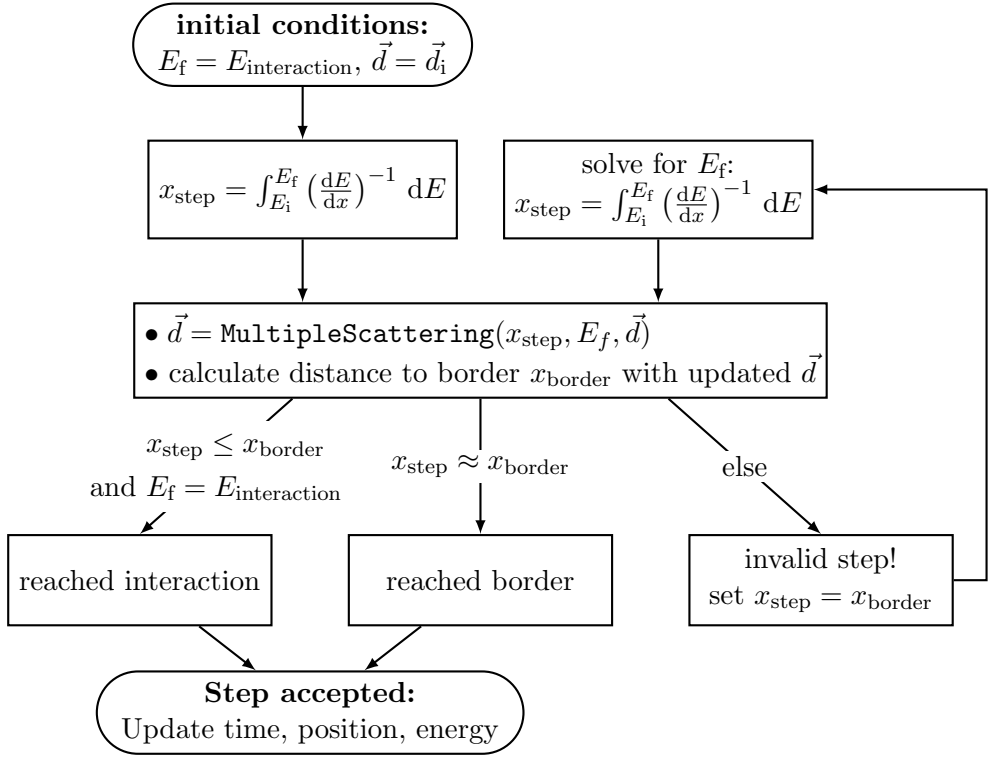
\begin{figure}
\begin{tikzpicture}[font=\small,thick,tips=proper]

\node[draw,
	rounded rectangle,
	align=center,
    minimum width=3.5cm,
    minimum height=1cm
] (block1) {\textbf{initial conditions:} \\ $E_\text{f} = E_\text{interaction}$, $\vec{d} = \vec{d_\text{i}}$};

\node[draw,
	below=0.7cm of block1,
	align=center,
    minimum width=3.5cm,
    minimum height=1.3cm
] (block2) {$x_\text{step} = \int_{E_\text{i}}^{E_\text{f}} \left( \frac{\mathrm{d}E}{\mathrm{d}x} \right)^{-1} \, \mathrm{d}E $};

\node[draw,
	right=0.6cm of block2,
	align=center,
    minimum width=3.5cm,
    minimum height=1.3cm
] (block8) {solve for $E_\text{f}$: \\$x_\text{step} = \int_{E_\text{i}}^{E_\text{f}} \left( \frac{\mathrm{d}E}{\mathrm{d}x} \right)^{-1} \, \mathrm{d}E $};

\node[draw,
	below=1.5cm of $(block2)!0.5!(block8)$,
	align=left,
    minimum width=3.5cm,
    minimum height=1cm
] (block3) {$\bullet \; \vec{d} = \texttt{MultipleScattering}(x_\text{step}, E_f, \vec{d})$ \\ $\bullet$ calculate distance to border $x_\text{border}$ with updated $\vec{d}$};
 
\node[draw,
	below=1.8cm of block3,
	align=center,
    minimum width=3.5cm,
    minimum height=1.2cm
] (block4) {reached border};
 
\node[draw,
	left=of block4,
	align=center,
    minimum width=3.5cm,
    minimum height=1.2cm
] (block5) {reached interaction}; 

\node[draw,
	right=of block4,
	align=center,
    minimum width=3.5cm,
    minimum height=1.2cm
] (block6) {invalid step!\\set $x_\text{step} = x_\text{border}$}; 

\node[draw,
	below=1.2cm of $(block4)!0.5!(block5)$,
	rounded rectangle,
	align=center,
    minimum width=3.5cm,
    minimum height=1cm
] (block7) {\textbf{Step accepted:}\\Update time, position, energy}; 


\draw[-latex] (block1) edge (block2);
\draw[-latex] (block2.south) edge (block2.south |- block3.north);

\draw[-latex] (block3) edge node[pos=0.5,fill=white,inner sep=2pt]{$x_\text{step} \approx x_\text{border} $} (block4.north);
\draw[-latex] ([xshift=-2.5cm] block3.south) edge node[pos=0.5,fill=white,inner sep=2pt, align=left]{$\begin{aligned} &x_\text{step} \leq x_\text{border} \\\text{and } &E_\text{f} = E_\text{interaction} \end{aligned}$} (block5.north);
\draw[-latex] ([xshift=2.5cm] block3.south) edge node[pos=0.5,fill=white,inner sep=2pt]{else} (block6.north);

\draw[-latex] (block4) edge (block7);
\draw[-latex] (block5) edge (block7);

\draw[-latex] (block8.south) edge (block8.south |- block3.north);

\draw [-latex] (block6.east) -| ([xshift=2.3cm] block8.east) -- (block8.east);

\end{tikzpicture}
\caption{Flowchart explaining the \texttt{AdvanceParticle} method of the \texttt{Propagator} class, handling multiple scattering in combination with border transitions. Note that the flowchart shows a simplified version of the algorithm, neglecting special cases such as limitations due to maximal propagation distances or minimal particle energies given by the user (see Section \ref{sec:termination_conditions}).}
\label{fig:border_flowchart}
\end{figure}

Therefore, an iterative approach to determine a valid combination of propagation step length and multiple scattering angles has been implemented as illustrated in a simplified version in Figure \ref{fig:border_flowchart}.
Instead of cutting the propagation step according to the updated distance to the border transition, the algorithm recalculates the multiple scattering translation using the updated distance.
This calculation is repeated until the difference between the propagation step and the distance to the border transition is within a given precision (by default \SI{e-3}{\centi\meter}).
In most cases, this algorithm converges after a few iterations.

\section{Summary} \label{sec:summary}

PROPOSAL is now a complete electromagnetic propagation module for high-energy charged leptons and photons, providing state-of-the-art parametrizations to sample energy losses and deflections.
New parametrizations and interaction types have been implemented to enable the propagation of electrons, positrons, and high-energy photons.
Instead of simply calculating the energy losses of the propagating particle, the option to calculate individual secondary particles, which can be further propagated, is now available.
Rare interaction processes have been implemented to reduce the uncertainty of muon and tau propagation.
Furthermore, model uncertainties concerning the photonuclear interaction can be better quantified due to a larger selection of interaction parameterizations.
Next to the physics-related improvements, numerous technical improvements have been made, ranging from more advanced interpolation methods, bug fixes, changes in the propagation routine and output, and an improved installation process, which makes PROPOSAL more user-friendly and flexible to use.

\section*{Acknowledgments}

This work was supported by the \emph{Deutsche Forschungsgemeinschaft} (Collaborative Research Center SFB 1491), the \emph{Bundesministerium für Bildung und Forschung}, and the \emph{Lamarr Institute}.
The authors thank the CORSIKA~8 collaboration for useful discussions.
We are grateful to Maximilian Linhoff for his technical expertise and support in maintaining PROPOSAL.

\appendix

\section{Rare muon interactions}
\subsection{Muon pair production by high-energy muons} \label{sec:mupair_xsection}
The parametrization for muon pair production implemented in PROPOSAL is taken
from \cite{Kelner2000mupair}. The cross section is given by
\begin{equation}
    \frac{\dif\sigma (E, v, \rho)}{\dif v\ \dif\rho} = \frac{2}{3 \pi} \left(Z \alpha r_e \frac{m_e}{m_\mu}
    \right)^2 \frac{1 - v}{v} \Phi(v, \rho) \ln X\,,
\end{equation}
with the relative energy loss $v = (E_{\mu^+} + E_{\mu^-}) / E$, the asymmetry parameter $\rho = (E_{\mu^+} - E_{\mu^-}) / (E_{\mu^+} + E_{\mu^-})$, and the definition
\begin{equation}
  \begin{aligned}
  \Phi(v, \rho) &= [(2 + \rho^2) (1 + \beta) + \xi (3 + \rho^2)] \ln \left(1 + \frac{1}{\xi}\right) \\
    & \quad + \left[(1 + \rho^2) \left(1 + \frac{3}{2} \beta\right) - \frac{1}{\xi}
    (1 + 2 \beta) (1 - \rho^2)\right] \ln (1 + \xi) \\
    &\quad - 1 - 3 \rho^2 + \beta (1 - 2 \rho^2)
  \end{aligned}
\end{equation}
with 
\begin{align}
  \beta &= v^2/[2 (1 - v)], & X &= 1 + U(E, v, \rho) - U(E, v, \rho_\text{max}), \\
  \xi &= v^2 (1 - \rho^2)/[4 (1 - v)],& \rho_\text{max} &= 1 - 2 m_\mu/(v E),
\end{align}
with
\begin{equation}
  U(E, v, \rho) = \frac{\num{0.65} A^{\num{-0.27}} B Z^{-1/3} m_\mu/m_e}
  {1 + \frac{2 \sqrt{e} m_\mu^2 B Z^{-1/3} (1 + \xi) (1 + Y)}{m_e E v
  (1 - \rho^2)}},
\end{equation}
and 
\begin{align}
  Y &= 12 \sqrt{m_\mu/E},& B &= 183.
\end{align}

\subsection{Weak Interaction} \label{sec:weak_xsection}
For the weak interaction of a charged lepton with a nucleus, the cross sections for neutrinos interacting with a nucleus can be used due to crossing symmetry.
Thereby, the neutrino cross section calculated by \cite{CSMS11NuXsection}, data of the HERA experiment\footnote{For the parton distribution functions, HERAPDF1.5 \cite{Aaron09herapdf10, CooperSarkar10herapdf15} was used.}, were interpolated for an energy range of \SIrange{10}{e12}{GeV}.
The differential cross sections were tabulated for the proton and neutron cross section.
The resulting cross section for PROPOSAL is averaged according to the number of individual nucleons in the nucleus
\begin{align}
    \frac{\dif \sigma}{\dif y} = \frac{Z}{A} \frac{\dif \sigma_\text{p}}{\dif y} + \left(1 - \frac{Z}{A} \right) \frac{\dif \sigma_\text{n}}{\dif y} .
\end{align}

Compared to most other interactions implemented in PROPOSAL, this purely stochastic interaction has no contribution to the continuous energy loss, and the cross section is differential in the Bjorken $y$, which describes the relative energy transferred to the nucleus.
The limits for the Bjorken $y$ in this parametrization are
\begin{align}
    y_{\text{min}} = \frac{Q_{\text{min}}^2}{E (m_\text{p} + m_\text{n}) + \left( \frac{m_\text{p} + m_\text{n}}{2} \right)^2}
    \qquad \text{and} \qquad
    y_{\text{max}} = 1 ,
\end{align}
where the neutrino mass is neglected for the upper limit, and $Q_{\text{min}}^2$ is set to \SI{1}{GeV^2}.
\section{New Photonuclear Parametrizations} \label{sec:newphotonucl_params}

The \texttt{AbtFT} parametrization \cite{abt2017photonucl} is a refit of the ALLM
parametrization \cite{ALLM91} to the combined HERA data \cite{HERA_combined} in
addition to fixed-target data \cite{E665,NMC,BCDMS}. It uses the same formulae
as \cite{ALLM91}, but with the parameters given in Table~\ref{tab:ALLM_FT}.
\begin{table}
  \begin{center}
  \begin{tabular}{cc|cc}
  \hline \hline
  Parameter & Value & Parameter & Value \\
  \hline
  $a_1^{\mathbb P}$ & \num{-0.075} & $a_1^{\mathbb R}$ & \num{0.882} \\
  $a_2^{\mathbb P}$ & \num{-0.470} & $a_2^{\mathbb R}$ & \num{0.082} \\
  $a_3^{\mathbb P}$ & \num{9.2}    & $a_3^{\mathbb R}$ & \num{-8.5}  \\
  \hline
  $b_1^{\mathbb P}$ & \num{-0.477} & $b_1^{\mathbb R}$ & \num{0.339} \\
  $b_2^{\mathbb P}$ & \num{54.0}   & $b_2^{\mathbb R}$ & \num{3.38} \\
  $b_3^{\mathbb P}$ & \num{0.073}  & $b_3^{\mathbb R}$ & \num{1.07} \\
  \hline
  $c_1^{\mathbb P}$ & \num{0.356}  & $c_1^{\mathbb R}$ & \num{-0.636} \\
  $c_2^{\mathbb P}$ & \num{0.171}  & $c_2^{\mathbb R}$ & \num{3,37} \\
  $c_3^{\mathbb P}$ & \num{18.6}   & $c_3^{\mathbb R}$ & \num{-0.660} \\
  \hline
  $m_0^2$ & \SI{0.338}{GeV^2}      & $m_\mathbb{R}^2$ & \SI{0.838}{GeV^2} \\
  $\Lambda^2$ & \SI{4.4e-9}{GeV^2} & $m_\mathbb{P}^2$ & \SI{50.8}{GeV^2} \\
  $Q_0^2$ & \SI{1.87e-5}{GeV^2}    \\
  \hline \hline
  \end{tabular}
  \end{center}
  \caption{Values of the parameters of the ALLM parametrization from the
    HHT-ALLM-FT fit in \cite{abt2017photonucl}.}
  \label{tab:ALLM_FT}
\end{table}

The \texttt{BlockDurandHa} parametrization was developed in \cite{block2014photonucl}
based on the idea of a saturated Froissart bound. In this model, the structure
function $F_2$ is parametrized as
\begin{multline}
  F_2(x, Q^2) = D(Q^2) (1 - x)^n \\
  \times \left[C(Q^2) + A(Q^2) \ln \left(\frac{Q^2/x}{Q^2 + \mu^2}\right) + B(Q^2)
    \ln^2 \left(\frac{Q^2/x}{Q^2 + \mu^2}\right) \right],
\end{multline}
with the squared four-momentum transfer $Q^2$ and the Bjorken scaling variable $x$, where
\begin{align}
  A(Q^2) &= a_0 + a_1 \ln \left(1 + \frac{Q^2}{\mu^2}\right)
    + a_2 \ln^2 \left(1 + \frac{Q^2}{\mu^2}\right), \\
  B(Q^2) &= b_0 + b_1 \ln \left(1 + \frac{Q^2}{\mu^2}\right)
    + b_2 \ln^2 \left(1 + \frac{Q^2}{\mu^2}\right), \\
  C(Q^2) &= c_0 + c_1 \ln \left(1 + \frac{Q^2}{\mu^2}\right), \\
  D(Q^2) &= \frac{Q^2 (Q^2 + \lambda M^2)}{(Q^2 + M^2)^2}.
\end{align}
The values of the parameters determined from the real photoabsorption
cross section parametrization by \cite{BlockHalzen} and combined HERA
data \cite{Aaron09herapdf10} are given in Table~\ref{tab:BlockDurandHa}.
\begin{table}
  \begin{center}
  \begin{tabular}{cc|cc}
    \hline \hline
    Parameter & Value & Parameter & Value \\
    \hline
    $a_0$ & \num{8.205e-4}  & $c_0$ & \num{0.255}     \\
    $a_1$ & \num{-5.148e-2} & $c_1$ & \num{1.475e-1}  \\
    $a_2$ & \num{-4.725e-3} & $n$ & \num{11.49} \\
    $b_0$ & \num{2.217e-3}  & $\lambda$ & \num{2.430} \\
    $b_1$ & \num{1.244e-2}  & $M^2$ & \SI{0.753}{GeV^2} \\
    $b_2$ & \num{5.958e-4}  & $\mu^2$ & \SI{2.82}{GeV^2} \\
    \hline \hline
  \end{tabular}
  \end{center}
  \caption{Fit values of the \texttt{BlockDurandHa} parametrization from
    \cite{block2014photonucl}.}
  \label{tab:BlockDurandHa}
\end{table}

\section{Deflection Parametrizations} \label{sec:deflection_params}
The following parametrizations by Van Ginneken \cite{vanginneken1986} have been validated for energies 
$\SI{3}{\giga\electronvolt}<E<\SI{30}{\tera\electronvolt}$.
A relative energy loss $\nu_{\mathrm{loss}}$ is introduced as 
\begin{equation}
    \nu_{\mathrm{loss}} = \frac{E_{\mathrm{loss}}}{E}\,,
\end{equation}
with the absolute the energy loss $E_{\mathrm{loss}}$, the initial muon energy $E$, and 
\begin{equation}
    \nu = (E - E^\prime) / (E - m_\mu)\,, 
\end{equation}
where $m_\mu$ is the muon mass, and $E^\prime$ the muon energy after the interaction.
For the root mean squared distributions $\langle \theta^2\rangle^{1/2}$ mentioned afterwards,  
the polar angle can be sampled from an exponential distribution as
\begin{equation}
    \theta = \sqrt{\lambda \cdot \exp{(-\lambda x)}}
\end{equation}
with $\lambda = \langle \theta^2\rangle$.

\subsection{Bremsstrahlung}

\subsubsection{Deflection parametrization of \textsc{Geant4}, Tsai}\label{sec:defl_brems_geant4}
In \textsc{Geant4} \cite{GEANT4}, angular distributions in $\theta_{\mu^\prime}$ for the outgoing muon $\mu^\prime$ and 
$\theta_\gamma$ for the photon $\gamma$
can be sampled via
\begin{equation}
    \theta_\gamma = \frac{m_\mu}{E}\,r\,, \qquad  \theta_{\mu^\prime} = \frac{E_{\mathrm{loss}}}{E^\prime}\,\theta_\gamma\,.
    \label{eqn:brems_theta_gamma_mu}
\end{equation}
The variable $r$ is defined as 
\begin{equation}
    r = \sqrt{\frac{a}{1-a}} \quad \text{with} \quad a = \xi \,\frac{r_{\mathrm{max}}^2}{1 + r_{\mathrm{max}}^2}\,.
\end{equation}
Here, $\xi$ is a random number uniformly distributed between $0$ and $1$ and 
the maximum $r_{\mathrm{max}}$ can be found as 
\begin{equation}
    r_{\mathrm{max}} = \min{\left(1, \frac{E^\prime}{E_{\mathrm{loss}}}\right)} \cdot \frac{E}{m_\mu}\,\theta^*\,.
\end{equation}
In this parametrization, the small-angle, ultrarelativistic approximation $E \gg m_\mu$ is
assumed which provides an accuracy of $\sim\!\SI{20}{\percent}$ at 
angles of $\theta \leq \theta^* \approx 1$.
The LPM effect leads to a suppression at energies $E \leq \SI{e20}{\electronvolt}$
which serves as an upper limit.
Due to rising of 
Ter-Mikaelyan at lower energy transfers, a lower limit is set to 
$\nu{_\mathrm{loss}} \geq \num{e-6}$, but bremsstrahlung is subdominant at such low 
$\nu_{\mathrm{loss}}$ \cite{Polityko_2002}. 

\subsubsection{Deflection parametrization from Van Ginneken}\label{sec:defl_brems_ginneken}
Based on the Mo-Tsai differential cross section 
$\mathrm{d}\sigma\,/\,\mathrm{d}p\,\mathrm{d}\Omega$ with momentum $p$
and solid angle $\Omega$ of the particle after photon emission, 
a fit of the muon deflection angle is derived by Van Ginneken \cite{vanginneken1986}. 
The RMS angle of an incoming muon of energy $E$ 
scattering 
at a nucleus
with atomic number $Z$ is 
found as 
\begin{equation}
    \langle\theta^2\rangle^{1/2} = 
    \begin{cases}
      \mathrm{max}\left[\mathrm{min}(k_1 \nu^{1/2}, k_2), k_3 \nu\right] & ,\nu \leq 0.5 \label{eqn:brems_param_vG_lower_05}\\
      k_4\nu^{1+n} (1 - \nu)^{-n} & ,\langle\theta^2\rangle^{1/2} < 0.2, \nu>0.5 \\
      k_5 (1- \nu)^{-1/2} & ,\langle\theta^2\rangle^{1/2} \geq 0.2, \nu > 0.5\,.
    \end{cases}
  \end{equation} 
The additional parameters for muons are defined as
\begin{align*}
    k_1 &= 0.092E^{-1/3}, & k_2 &= 0.052E^{-1}Z^{-1/4}, \\
    k_3 &= 0.22E^{-0.92}, & k_4 &= 0.26E^{-0.91},\\
    k_5 &= k_4 \nu_{\mathrm{g}}^{1+n} (1-\nu_{\mathrm{g}})^{0.5-n}, & n &= 0.81E^{0.5} / (E^{0.5} + 1.8),
\end{align*}
where $\nu_{\mathrm{g}}$ follows with continuity at $\langle \theta^2\rangle^{1/2} = 0.2\,, \nu > 0.5$ .

\subsection{Electron-positron Pair Production}

\subsubsection{Deflection parametrization from Van Ginneken}\label{sec:defl_epair_ginneken}
A differential cross section for pair production is derived by Kelner \cite{kelner67}. 
A fit for the RMS muon deflection is performed by Van Ginneken as  
\begin{equation}
    \langle \theta^2\rangle^{1/2} = (2.3 + \ln{(E)})\, \frac{1}{E\nu^2}\, (1-\nu)^n \,\left(\nu - \frac{2m_{e}}{E}\right)^2 \cdot \min{[\,f(\nu, E, a, b, c, d), e\,]}\,,
    \label{eqn:pairprod_rms_theta}
\end{equation} 
with
\begin{equation}
    f = a\,\nu^{1/4} (1 + b\,E) + \frac{c\,\nu}{\nu + d}.
\end{equation}
For muons, the parameters in the parametrization are defined as
\begin{align*}
    n &= -1, & a &= \num{8.9e-4}, & b &= \num{1.5e-5}, \\
    c &= 0.0032, & d &= 1.0, & e &= 0.1.
\end{align*}
A dependence on the charge number $Z$ is neglected.
A minimum energy transfer 
$E_{\mathrm{loss,\min}} = E - E_{\mathrm{max}}^\prime$ follows as 
\begin{alignat}{3}
    & \qquad\nu_{\mathrm{min}} &&= \frac{2m_{e}}{E} \\
    \Leftrightarrow \quad& \qquad E_{\mathrm{loss,\min}} &&= 2m_e \left(1 - \frac{m}{E}\right) < 2m_{e}\,.
\end{alignat}
In practical use the limit is set to $\nu_{\mathrm{min,p}} \approx 4m_{e}/E$ 
\cite{pair_prod_lim} due to the relativistic approximations in the 
cross sections used. 

\subsection{Ionization}

\subsubsection{Deflection parametrization by kinematics}\label{sec:defl_ioniz_naive}
A calculation via four-momentum conservation is performed to determine the resulting deflection of the incoming muon.
The interaction is expressed by 
\begin{equation}
    P_\mu + P_{e} = P_{\mu^\prime} + P_{e^\prime}\,,
    \label{eqn:ionization_four_momentum}
\end{equation}
where $P_i$ and $P_{i^\prime}$ describe the incoming and outgoing four-momenta of the interacting muon and electron. To calculate the resulting deflection $\theta$ of the incoming muon, the following assumptions are used:
\begin{itemize}
    \item $E_{e^\prime} = E_\mu - E_{\mu^\prime} + E_{e}:$ The resulting electron energy depends on the electron energy before the interaction in addition to the total energy loss of the muon due to energy conservation.
    \item $\vec{p}_{e}^{\,2} \approx 0 \Rightarrow E_{e} \approx m_{e}:$ The atomic electron rests before the interaction.
\end{itemize}
By squaring and transforming Equation~\eqref{eqn:ionization_four_momentum} using the previously mentioned assumptions, the muon deflection can be found 
as 
\begin{equation}
    \theta = \arccos{\left(\frac{(E_\mu + m_{e}) E_{\mu^\prime} - E_\mu m_{e} - m_\mu^2}{|\vec{p}_\mu| \cdot |\vec{p}_{\mu^\prime}|}\right)}\,.
    \label{eqn:ioniz_formula}
\end{equation}

\subsection{Photonuclear Interaction}

\subsubsection{Deflection parametrization of \textsc{Geant4}}\label{sec:defl_photonuclear_geant4}
To sample muon deflection angles due to photonuclear interaction, 
in \textsc{Geant4} \cite{GEANT4} the differential cross section 
is written as 
\begin{equation}
    \frac{\mathrm{d}\sigma}{\mathrm{d}t} \sim \frac{\left(1 - \frac{t}{t_{\mathrm{max}}}\right)}{t\left(1 + \frac{t}{E_{\mathrm{loss}}^2}\right)\left(1 + \frac{t}{m_0^2}\right)} \left[(1-\nu_{\mathrm{loss}})\left(1 - \frac{t_{\mathrm{min}}}{t}\right) + \frac{\nu_{\mathrm{loss}}^2}{2}\right]\,, \label{eqn:photonuclear_cross}
\end{equation}
with the assumption of four-momentum transfers $Q^2\leq\SI{3}{\giga\electronvolt\squared}$ \cite{osti_7308048, borog_petrukhin_cross}. 
Here, the definition $t = Q^2 = 2 (EE^\prime - PP^\prime \cos{\theta} - m_\mu^2)$ is used with the parameters
\begin{align*}
    t_{\mathrm{min}} &= \frac{(m_\mu \,\nu_{\mathrm{loss}})^2}{1-\nu_{\mathrm{loss}}}, & \nu_{\mathrm{loss}} &= \frac{E_{\mathrm{loss}}}{E},\\ 
    t_{\mathrm{max}} &= 2ME_{\mathrm{loss}}, & E_{\mathrm{loss}} &= E - E^\prime,
\end{align*}
with the nucleon (here proton) mass $M$, and $m_0^2 \equiv \Lambda^2 \approx \SI{0.4}{\giga\electronvolt\squared}$, which is a phenomenological parameter determining the behavior of the inelastic form factor. 
Weak interaction factors are neglected on $t$.
Equation~\eqref{eqn:photonuclear_cross} can be written as 
\begin{equation}
    \sigma(t) \sim f(t)\,g(t)\,,
\end{equation}
with two substitutions
\begin{align}
    \begin{split}
    f(t) &= \frac{1}{t\left(1 + \frac{t}{t_1}\right)}\,, \\
    g(t) &= \frac{1 - \frac{t}{t_{\mathrm{max}}}}{1 + \frac{t}{t_2}} \cdot \frac{(1 - \nu_{\mathrm{loss}})\left(1 - \frac{t_{\mathrm{min}}}{t}\right) + \frac{\nu_{\mathrm{loss}}^2}{2}}{(1-\nu_{\mathrm{loss}}) + \frac{\nu_{\mathrm{loss}}^2}{2}}\,,
    \end{split}
    \label{eqn:photonuclear_formula}
\end{align}
and 
\begin{equation}
    t_1 = \min{(E_{\mathrm{loss}}^2, m_0^2)}\,, \qquad t_2 = \max{(E_{\mathrm{loss}}^2, m_0^2)}\,.
\end{equation}
Solving Equations~\eqref{eqn:photonuclear_formula} analytically leads to 
\begin{equation}
    t_P = \frac{t_{\mathrm{max}}t_1}{(t_{\mathrm{max}} + t_1) \left[\frac{t_{\mathrm{max}}(t_{\mathrm{min}} + t_1)}{t_{\mathrm{min}}(t_{\mathrm{max}} + t_1)}\right]^\xi - t_{\mathrm{max}}}\,,
    \label{eqn:t_p_nuclint_param}
\end{equation}
with a random number $\xi \in (0, 1)$ and acceptance of probability $g(t)$. 
From this, the polar muon deflection angle $\theta$ can be derived as 
\cite{GEANT4}
\begin{equation}
    \theta = 2 \cdot \arcsin{\left(\sqrt{\frac{t_P - t_{\mathrm{min}}}{4(EE^\prime - m_\mu^2) - 2\,t_{\mathrm{min}}}}\,\right)}\,.
    \label{eqn:nucl_int_theta_G4}
\end{equation}
A minimum energy loss of $\SI{200}{\mega\electronvolt}$ and a maximum 
energy transfer of $E - M / 2$ with $M$ as the nucleon (proton)
mass is recommended.

\subsubsection{Deflection parametrization from Van Ginneken}\label{sec:defl_photonuclear_ginneken}
Using a lepto-production scaling formula from \cite{lepto_production}, 
Van Ginneken has fit the RMS angle of the outgoing muon as 
\begin{equation}
    \langle \theta^2\rangle^{1/2} = \frac{0.39}{E(1 - \nu)} (E^{1/2}\nu(1-\nu))^{0.17} \cdot \left(1 - \frac{0.135}{E\nu}\right)\,.
    \label{eqn:nucl_int_param_vG}
\end{equation}
A minimum energy transfer 
results to about $\SI{136}{\mega\electronvolt}$ since this is the mass 
of the pion, which is the lightest hadron that can be produced
\cite{vanginneken1986}.

\begin{figure}
    \centering 
    \begin{subfigure}{.48\textwidth}
        \centering
        \includegraphics[width=\linewidth]{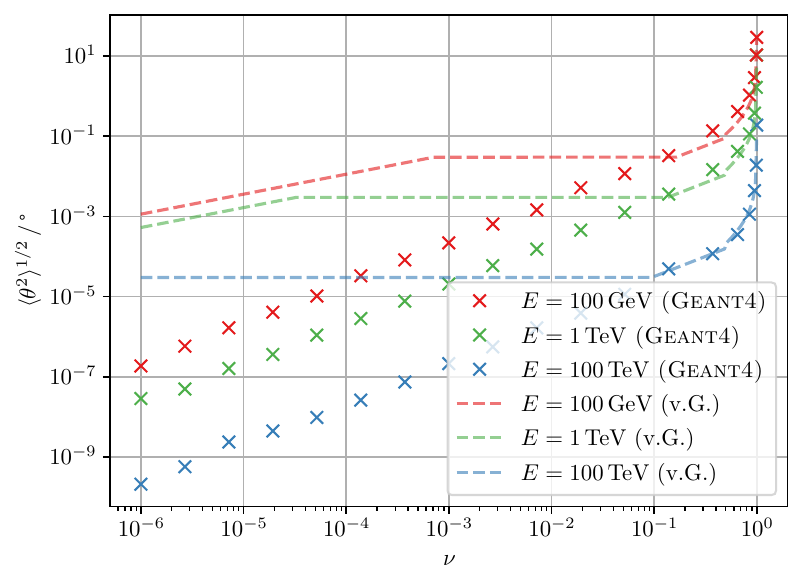}  
        \caption{Root mean squared deflections due to bremsstrahlung parametrized by \textsc{Geant4} are lower than those determined by Van Ginneken for 
        very small energy transfers. In \textsc{Geant4}, an approximation of bremsstrahlung in the Coulomb filed of an 
        unshielded nucleus is used, what leads to the non-appearance of the plateau. At high $\nu$, the agreement is better. The 
        atomic number is set to $Z = 1$.}
        \label{fig:defl_brems}
    \end{subfigure}
    \hfill
    \begin{subfigure}{.48\textwidth}
        \centering
        \includegraphics[width=\linewidth]{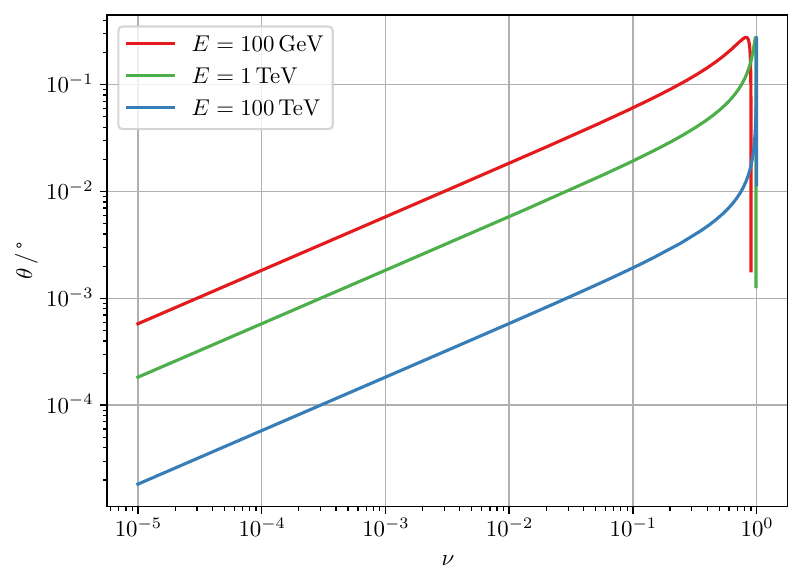}  
        \caption{Calculation of deflection due to ionization is shown. 
        For very high energy 
        transfers, the deflection approaches zero due to a vanishing transverse momentum. All distributions peak with $\theta = \SI{0.277}{\degree}$.
        Maximum energy transfers result to minimum final energies $E^\prime(E = \SI{100}{\giga\electronvolt}) > \SI{9.85}{\giga\electronvolt}$, 
        $E^\prime(E = \SI{1}{\tera\electronvolt}) > \SI{10.81}{\giga\electronvolt}$ and $E^\prime(E = \SI{100}{\tera\electronvolt}) 
        > \SI{10.92}{\giga\electronvolt}$.}
        \label{fig:defl_ioniz}
    \end{subfigure}
    \newline
    \begin{subfigure}{.48\textwidth}
        \centering
        \includegraphics[width=\linewidth]{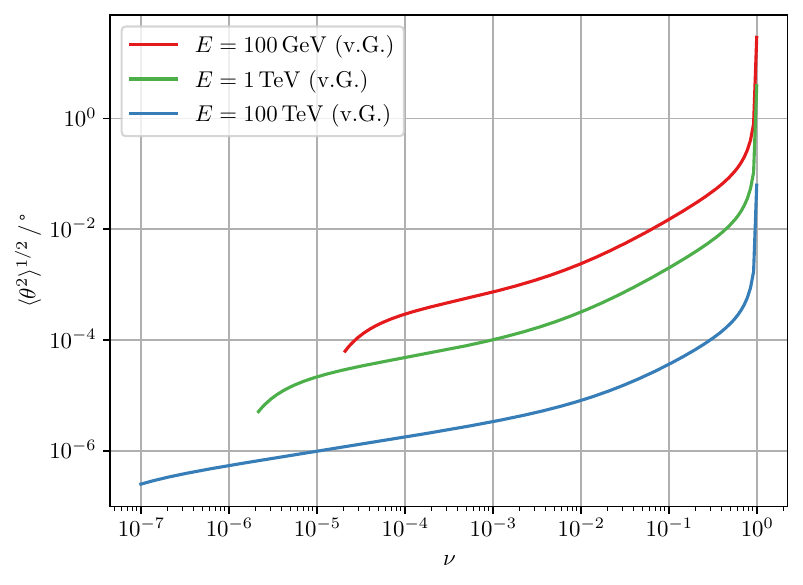}  
        \caption{Root mean squared deflections for electron pair production, parametrized by Van Ginneken, are shown. For very small 
        relative energy transfers $\nu$, the deflection increases sharply.}
        \label{fig:defl_epair}
    \end{subfigure}
    \hfill
    \begin{subfigure}{.48\textwidth}
        \centering
        \includegraphics[width=\linewidth]{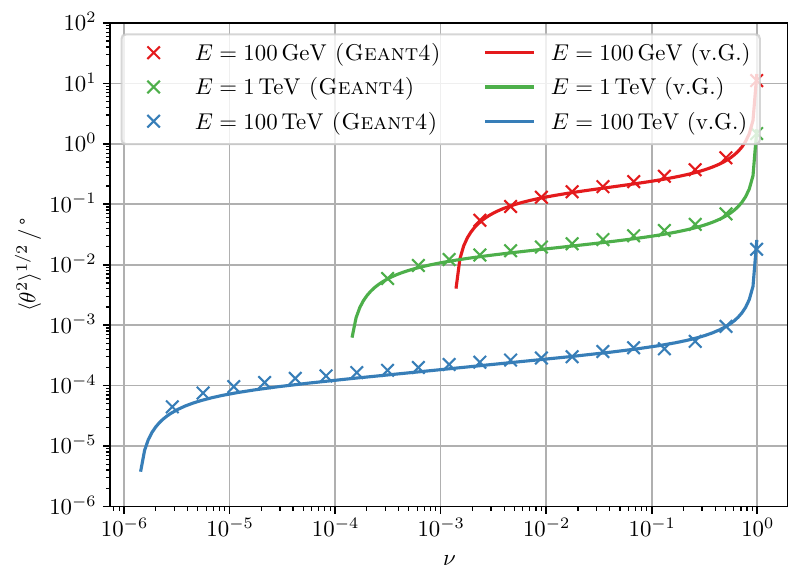}  
        \caption{Root mean squared deflections due to photonuclear interactions are parametrized by Van Ginneken 
        and \textsc{Geant4} with minimum energy transfers $\nu_{\mathrm{min,\,\textsc{Geant4}}} = \SI{200}{\mega\electronvolt}$ and 
        $\nu_{\mathrm{min,\,v.G.}} = \SI{136}{\mega\electronvolt}$.}
        \label{fig:defl_photonuclear}
    \end{subfigure}
    \caption{All parametrizations to describe deflections by stochastic interactions implemented in PROPOSAL 
    are presented for three different energies in dependence of the relative energy transfer $\nu = (E - E^\prime)/(E - m)$
    with the initial particle energy $E$, the remaining energy $E^\prime$ after the interaction and the muon 
    mass $m$. In each sampling, $\num{100000}$ angles are drawn. For bremsstrahlung and photonuclear interaction, there are two parametrizations available each by 
    Van Ginneken (v.G.) \cite{vanginneken1986} and \textsc{Geant4} \cite{GEANT4}. The lower the energy, the larger the deflection.
    Bigger energy transfers result in larger deflections.}
    \label{fig:all_deflections_rms}
\end{figure}

\section{Electron and Positron Interactions} \label{sec:ep_params}

\subsection{Ionization}
\label{sec:ionization}
Ionization describes the energy loss due to collisions with atomic electrons.
For energy transfers high compared to the binding energies of atomic electrons, the ionization process can be considered a scattering process on free electrons and, therefore, described by M{\o}ller and Bhabha scattering.
In this case, the differential cross section is given for $T \gg I$, with $T$ the kinetic energy of the secondary electron and $I$ the ionization energy of the target
atom, as
\begin{equation}
	\frac{d\sigma}{d\epsilon} = \frac{2 \pi r_e^2 Z}{\beta^2 (\gamma - 1)}
	\left[\frac{(\gamma - 1)^2}{\gamma^2} + \frac{1}{\epsilon} \left(\frac{1}
	{\epsilon} - \frac{2 \gamma - 1}{\gamma^2} \right) + \frac{1}{1 - \epsilon}
	\left(\frac{1}{1 - \epsilon} - \frac{2 \gamma - 1}{\gamma^2} \right) \right]
\end{equation}
for M{\o}ller scattering $e^- e^- \rightarrow e^- e^-$, and
\begin{equation}
	\frac{d\sigma}{d\epsilon} = \frac{2 \pi r_e^2 Z}{\gamma - 1} \left[\frac{1}
	{\beta^2 \epsilon^2} - \frac{B_1}{\epsilon} + B_2 - B_3 \epsilon + B_4
	\epsilon^2\right]
\end{equation}
for Bhabha scattering $e^+ e^- \rightarrow e^+ e^-$. Here, $\epsilon = T/(E -
m c^2)$ with $E$ the initial electron/positron energy.
The kinematic limits of $\epsilon$ are given as
\begin{align}
	\epsilon_0 &= \frac{T_\text{cut}}{E - m c^2} \leq \epsilon \leq \frac{1}{2},
	\text{ for } e^- e^-,\\
	\epsilon_0 &= \frac{T_\text{cut}}{E - m c^2} \leq \epsilon \leq 1, \text{ for }
	e^+ e^-.
\end{align}
Here $T_\text{cut}$ denotes a threshold energy.
The quantities $B_i$ are given by
\begin{align}
	B_1 &= 2 - y^2, & \gamma &= E/m c^2,\\
	B_2 &= (1 - 2 y) (3 + y^2), & \beta^2 &= 1 - 1/\gamma^2,\\
	B_3 &= (1 - 2 y)^2 + (1 - 2 y)^3, & y &= \frac{1}{\gamma + 1},\\
	B_4 &= (1 - 2 y)^3.
\end{align}

\begin{figure}
    \centering
    \includegraphics[width=0.8\textwidth]{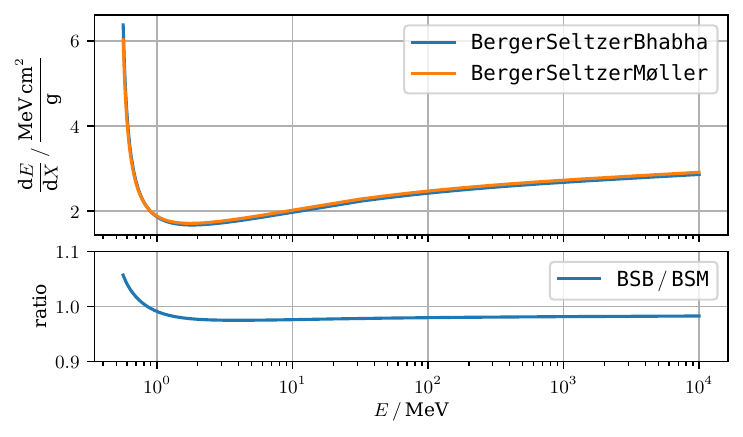}
    \caption{Continuous energy losses of electrons (M{\o}ller) and positrons (Bhabha) in air.}
    \label{fig:ioniz_cross}
\end{figure}

The electron binding energies need to be considered for small energy transfers, which are relevant for calculating continuous energy losses.
This continuous energy loss is described by the parametrization by Berger and Seltzer \cite{bergerseltzer}, which is used in PROPOSAL to describe continuous ionization losses for electrons and positrons.
It is given by
\begin{equation}
	\label{eqn:berger_seltzer}
	- \left(\frac{\mathrm{d}E}{\mathrm{d}X}\right) = \frac{2 \pi r_e^2 m_e}{\beta^2} \left[ \ln{\left( \frac{2 m_e (\tau + 2)}{I} \right) + F^{\pm}(\tau, \Delta) - \delta } \right],
\end{equation}
with the mean ionization energy $I$, and
\begin{align}
	\tau &= \gamma - 1, & \Delta = \text{min}\left( \frac{v_{\text{max}} E}{m_e}, \frac{v_{\text{cut}} E}{m_e} \right).
\end{align}
For the density correction $\delta$, the parametrization by Sternheimer \cite{PhysRevB.3.3681} is used.
The term $F^{\pm}(\tau, \Delta)$ is given by 
\begin{align}
	\begin{split}
	F^{-}(\tau, \Delta) &= -1 - \beta^2 + \ln\left( (\tau - \Delta) \Delta \right) + \frac{\tau}{\tau - \Delta} \\ &\quad+ \frac{1}{\gamma^2} \left[ \frac{\Delta^2}{2} + (2 \tau + 1) \ln\left( 1 - \frac{\Delta}{\tau} \right) \right]\\
	\end{split}
	\intertext{for electrons, and}
	\begin{split}
	F^{+}(\tau, \Delta) &= \ln\left(\tau \Delta \right) - \frac{\beta^2}{\tau} \biggl[ \tau + 2 \Delta - \frac{3 \Delta^2 y }{2} \\ &\quad- (\Delta - \frac{\Delta^2}{3}) y^2 - ( \frac{\Delta^2}{2} - \frac{\tau\Delta^3}{3} + \frac{\Delta^4}{4} ) y^3 \biggr]
	\end{split}
\end{align}
for positrons.
Figure \ref{fig:ioniz_cross} shows the average ionization energy loss of electrons and positrons.

\subsection{Bremsstrahlung}
\label{sec:bremsstrahlung}
For energies above $\SI{50}{\mega\electronvolt}$, the ultra-relativistic differential cross section, comparable to the Complete Screening parametrization, which has already been implemented in PROPOSAL, is given by \cite{RevModPhys.31.920, Hirayama:2005zm}
\begin{equation}
	\label{eqn:brems_high}
	\begin{split}
	\frac{d\sigma}{dv} &= \frac{Z (Z + \xi(Z)) r_e^2 \alpha}{v} \biggl[ (2 - 2 v + v^2) \left( \Phi_1(x) - \frac{4}{3} \ln(Z) - 4 f_c(Z) \right) \\ &\quad- \frac{2}{3} (1 - v) \left( \Phi_2(x) - \frac{4}{3} \ln(Z) - 4 f_c(Z) \right) \biggl],
	\end{split}
\end{equation}
with the relative energy loss $v$, and
\begin{align}
	x &= 136 Z^{-\sfrac{1}{3}} \frac{2 \delta}{m_e}, & \delta &= \frac{m_e^2 v}{2 E (1 - v)}.
\end{align}
The functions $\Phi_1(x)$, $\Phi_2(x)$ describe the screening effects and are given by
\begin{align}
	\Phi_1(x) &= 
	\begin{cases}
		20.867 - 3.242 x + 0.625 x^2 & \text{if $x \leq 1$},\\
		21.12 - 4.184 \ln(x + 0.952) & \text{if $x > 1$},\\
	\end{cases}\\
	\Phi_2(x) &= 
	\begin{cases}
		20.029 - 1.930 x - 0.086 x^2 & \text{if $x \leq 1$},\\
		21.12 - 4.184 \ln(x + 0.952) & \text{if $x > 1$},\\
	\end{cases}
\end{align}
which is an analytical approximation of the Thomas-Fermi form factors \cite{BUTCHER196015}.
Furthermore, $f_c(Z)$ describes the Coulomb correction and is approximated in an analytical form by \cite{PhysRev.93.788}
\begin{equation}
	f_c(Z) \approx a^2 \left[ \frac{1}{1 + a^2} + 0.20206 - 0.0369a^2 + 0.0083 a^4 - 0.002 a^6 \right]
\end{equation}
with $a = \alpha Z$.
The parameter 
\begin{equation}
	\xi(Z) = \frac{L_\text{rad}^{\prime}(Z)}{L_\text{rad}(Z) - f_c(Z)} 
\end{equation}
with the radiation logarithms \cite{tsai1974}
\begin{align}
	L_\text{rad}^{\prime} &=
	\begin{cases}
		\ln(1194 Z^{-\sfrac{2}{3}}) & \text{if $Z > 4$}, \\
		5.924 & \text{if $Z = 4$}, \\
		5.805 & \text{if $Z = 3$}, \\
		5.621 & \text{if $Z = 2$}, \\
		6.144 & \text{if $Z = 1$},
	\end{cases} \\
	L_\text{rad} &=
	\begin{cases}
		\ln(184.15 Z^{-\sfrac{1}{3}}) & \text{if $Z > 4$}, \\
		4.710 & \text{if $Z = 4$}, \\
		4.740 & \text{if $Z = 3$}, \\
		4.790 & \text{if $Z = 2$}, \\
		5.310 & \text{if $Z = 1$},
	\end{cases}
\end{align}
accounts for atomic electron effects.

For energies below $\SI{50}{\mega\electronvolt}$, the differential bremsstrahlung cross section is given by
\begin{equation}
	\label{eqn:brems_low}
	\begin{split}
	\frac{d\sigma}{dv} &= \frac{A^{\prime}(E, Z) Z (Z + \xi(Z)) r_e^2 \alpha}{v} \biggl[ (2 - 2 v + v^2) \left( \Phi_1(x) - \frac{4}{3} \ln(Z) \right) \\ &\quad- \frac{2}{3} (1 - v) \left( \Phi_2(x) - \frac{4}{3} \ln(Z) \right) \biggl]
	\end{split}	
\end{equation}
where a density correction factor $A^{\prime}(E,Z)$ has been introduced \cite{pirs}.
This factor $A^{\prime}(E,Z)$ rescales the differential cross section to agree with the empirical average energy losses per distance from \cite{icru37}.
It should be noted that this factor is only a normalization factor and is, therefore, not affecting the shape of the energy loss distribution.

\subsection{Electron-positron Pair Production}
\label{sec:electron_positron_pairproduction}
The cross section for the process of triplet pair production $e^\pm + Z \rightarrow
e^\pm + e^+ + e^- + Z$ was adapted from the cross section for muon-induced muon pair
production $\mu^\pm + Z \rightarrow \mu^\pm + \mu^+ + \mu^- + Z$ of \cite{Kelner2000mupair}, removing
the nuclear form factor correction, which for electrons is negligible in contrast
to muons. The cross section is parametrized by
\begin{equation}
  \frac{d\sigma}{dv\, d\rho} = \frac{2}{3 \pi} (Z \alpha r_e)^2 \frac{1 - v}{v}
  \Phi \ln X,
\end{equation}
where
\begin{align}
\begin{split}
  \Phi &= [(2 + \rho^2) (1 + \beta) + \xi (3 + \rho^2)]
    \ln \left(1 + \frac{1}{\xi}\right) \\
  &+ \left[(1 + \rho^2) \left(1 + \frac{3}{2} \beta\right)
    - \frac{1 + 2 \beta}{\xi} (1 - \rho^2) \right] \ln (1 + \xi) \\
  &- 1 - 3 \rho^2 + \beta (1 - 2 \rho^2),
\end{split} \\
  X &= 1 + U(E, v, \rho) - U(E, v, \rho_\text{max}), \\
  U(E, v, \rho) &= \frac{B Z^{-1/3}}{1 + \frac{2 \sqrt{e} m B Z^{-1/3}
    (1 + \xi) (1 + Y)}{E v (1 - \rho^2)}}
\end{align}
with
\begin{align*}
  \beta &= \frac{v^2}{2 (1 - v)}, & \xi &= \frac{v^2 (1 - \rho^2)}{4 (1 - v)},\\
  \rho_\text{max} &= 1 - \frac{2 m}{E v}, & Y &= 12 \sqrt{m/E}.
\end{align*}

\subsection{Annihilation}
\label{sec:annihilation}
The total cross section for annihilation of a positron with Lorentz factor
$\gamma$ on an atomic electron at rest is given by \cite{heitler1954, GEANT4}
\begin{equation}
	\sigma(\gamma) = \frac{\pi r_e^2}{\gamma + 1} \left[\frac{\gamma^2 + 4 \gamma
	+ 1}{\gamma^2 - 1} \ln (\gamma + \sqrt{\gamma^2 - 1}) - \frac{\gamma + 3}
	{\sqrt{\gamma^2 - 1}}\right].
\end{equation}
The differential cross section for annihilation into two photons is given by
\begin{equation}
	\frac{d\sigma}{d\epsilon} = \frac{\pi r_e^2}{\gamma - 1} \frac{1}{\epsilon}
	\left[1 + \frac{2 \gamma}{(\gamma + 1)^2} - \epsilon - \frac{1}
	{(\gamma + 1)^2} \frac{1}{\epsilon}\right],
\end{equation}
where $\epsilon$ is the dimensionless photon energy $\epsilon = E_\gamma/m c^2$,
which has the limits
\begin{equation}
	\epsilon_\text{min, max} = \frac{1}{2} \left[1 \pm \sqrt{\frac{\gamma - 1}
	{\gamma + 1}}\right].
\end{equation}
The final state kinematics can be determined from energy-momentum conservation.
The scattering angle $\theta$ between the incident positron and the photon with
energy $\epsilon$ is given by
\begin{equation}
	\cos\theta = \frac{\epsilon (\gamma + 1) - 1}{\epsilon \sqrt{\gamma^2 - 1}}.
\end{equation}

\section{Photon Interactions} 
\label{sec:gamma_params}

\subsection{Compton Scattering}
\label{sec:compton}
The differential Klein-Nishina cross section for scattering of a photon with
energy $\omega_0$ to a final state with photon energy
\begin{equation}
	\omega_1 = \frac{\omega_0}{1 + (\omega_0/m) (1 - \cos \theta)},
	\label{eq:compton_relation}
\end{equation}
where $\theta$ is the scattering angle, is given by \cite{klein_nishina}
\begin{equation}
	\frac{d\sigma}{d\Omega} = \frac{Z r_e^2}{2} \left(\frac{\omega_0}{\omega_1}
	\right)^2 \left[4 \left(\frac{m}{2 \omega_1} - \frac{m}{2 \omega_0}\right)^2
	- 4 \left(\frac{m}{2 \omega_1} - \frac{m}{2 \omega_0}\right)
	- \left(\frac{\omega_1}{\omega_0} + \frac{\omega_0}{\omega_1}\right)\right].
\end{equation}
With the Compton relation \eqref{eq:compton_relation}, the cross section
differential in $\omega_1$ is given by
\begin{equation}
	\frac{d\sigma}{d\omega_1} = Z \pi r_e^2 \frac{\omega_0^2 m}{\omega_1^4}
	\left[4 \left(\frac{m}{2 \omega_1} - \frac{m}{2 \omega_0}\right)^2
	- 4 \left(\frac{m}{2 \omega_1} - \frac{m}{2 \omega_0}\right)
	- \left(\frac{\omega_1}{\omega_0} + \frac{\omega_0}{\omega_1}\right)\right].
\end{equation}
The total cross section can be calculated analytically as
\begin{equation}
	\sigma(\epsilon) = Z \frac{\pi r_e^2}{\epsilon^2} \left(4 + \frac{2 \epsilon^2
	(1 + \epsilon)}{(1 + 2 \epsilon)^2} + \frac{\epsilon^2 - 2 \epsilon - 2}
	{\epsilon} \ln (1 + 2 \epsilon) \right),
\end{equation}
where $\epsilon = E_\gamma/m_e c^2$ is the photon energy in the electron rest
frame in units of the electron rest mass.

\subsection{Electron-positron Pair Production}
\label{sec:photopairproduction}

\begin{figure}
    \centering
    \includegraphics[width=0.8\textwidth]{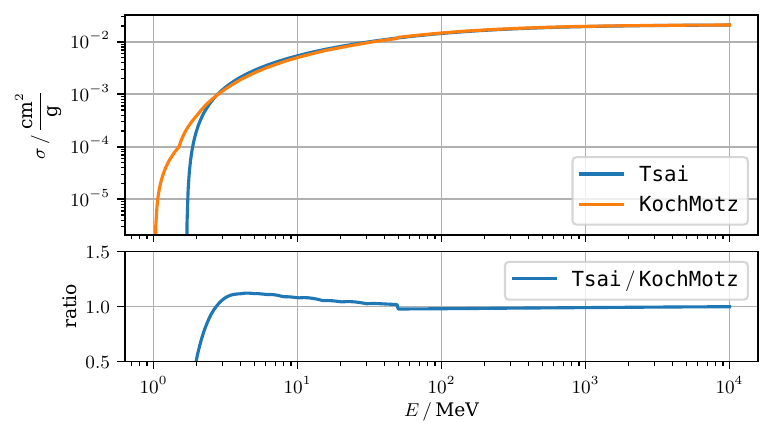}
    \caption{Total cross section of electron-positron pair production in air for the parametrizations according to \eqref{eqn:crosssection_pairproduction} (\texttt{Tsai}) and \eqref{eqn:crosssection_pairproduction_kochmotz} (\texttt{KochMotz}).}
    \label{fig:photopair_cross}
\end{figure}

In PROPOSAL, the process of electron-positron pair production can be described by two different parametrizations presented in this section.
Figure \ref{fig:photopair_cross} compares the two parametrizations.
Furthermore, the LPM suppression for this process has been implemented.

\subsubsection{Tsai parametrization}

The process of electron-positron pair production is described by quantum electrodynamics, and a differential cross section exact to order $\alpha^3$ is provided by \cite{tsai1974}.
However, evaluating this expression is too complicated for a direct implementation.
Based on \cite{tsai1974}, an approximated expression is given by
\begin{equation}
  \label{eqn:crosssection_pairproduction}
  \begin{split}
  \frac{d\sigma}{dx} &= \frac{\alpha r_e^2 x E_{\gamma}}{p} \biggl\{ \left(\frac{4}{3} x^2 - \frac{4}{3} x + 1 \right) \\ &\quad\times \biggl[ Z^2 \left(\varphi_1 - \frac{4}{3} \ln(Z) - 4f(z) \right) + Z \left( \psi_1 - \frac{8}{3} \ln(Z) \right) \biggr]  \\ &\quad- \frac{2}{3} x (1 - x) \biggl[ Z^2 \left(\varphi_1 - \varphi_2\right) + Z \left(\psi_1 - \psi_2\right) \biggr] \biggr\}
  \end{split}
\end{equation}
with $x = E_{-} / E_{\gamma}$, where $E_{\gamma}$ is the energy of the initial photon, $E_-$ the energy of the produced electron, and $p$ its absolute momentum.
The function $f(z)$ describes the Coulomb correction and is defined as \cite{tsai1974}
\begin{align}
  f(z) &= 1.202 z - 1.0369 z^2 + 1.008 \frac{z^3}{1+z}, & z &= \left(\frac{Z}{137}\right)^2.
\end{align}
While the approximate differential cross section ignores effects from nuclear form factors, which are only important for large production angles, the effects from the atomic form factors are described by the functions $\varphi_1$, $\varphi_2$ (elastic scattering part) and $\psi_1$, $\psi_2$ (inelastic scattering part).
The description used for the atomic form factors varies with $Z$, and the resulting expressions for $\varphi$ and $\psi$ are given in \cite{tsai1974}.

Since the photon must provide the rest mass of both the electron and the positron, the kinematic limits of the process are given by
\begin{align}
  E &\geq 2 m_e, & x_{\text{min}} &= \frac{m_e}{E_{\gamma}}, & x_{\text{max}} &= 1 - \frac{m_e}{E_{\gamma}}.
\end{align}

\subsubsection{Koch and Motz parametrization}
\label{sec:kochmotz}

The cross section of electron-positron pair production is given analogously to the bremsstrahlung cross section in \ref{sec:bremsstrahlung}.
Above \SI{50}{MeV}, the cross section is given by \cite{Hirayama:2005zm}
\begin{equation}
  \label{eqn:crosssection_pairproduction_kochmotz}
  \begin{split}
    \frac{d\sigma(Z, E, v)}{dx} &= Z (Z + \xi(Z)) r_0^2 \alpha
    \left\{(2 x^2 - 2 x + 1) \left[\Phi_1(\delta) - \frac{4}{3} \ln Z - 4 f_c(Z)
    \right] \right.\\
    &+ \left. \frac{2}{3} x (1 - x) \left[\Phi_2(\delta) - \frac{4}{3} \ln Z
    - 4 f_c(Z)\right]\right\},
  \end{split}
\end{equation}
where
\begin{equation}
  \begin{split}
    \delta &= \frac{136 m_e Z^{-1/3}}{E_\gamma x (1 - x)}.
  \end{split}
\end{equation}
The structure functions $\Phi_{1,2}$ are identical to the bremsstrahlung case.
Again, for energies below \SI{50}{MeV}, an empirical correction factor is used, with cross section data taken from \cite{storm_israel}.

\subsubsection{LPM effect}
\label{sec:lpm_photopair}

The suppression of the electron-positron pair production cross section due to the LPM effect has been implemented based on the parametrization given in \cite{RevModPhys.71.1501} and is defined as 
\begin{align}
  \label{eqn:lpm_photopair}
  \frac{\mathrm{d}\sigma_\text{LPM}}{\mathrm{d}x} &= \frac{\mathrm{d}\sigma}{\mathrm{d}x} \cdot \frac{\xi(s) / 3 \left(G(s) + 2 \left( x^2 + (1 - x)^2 \right) \phi(s) \right)}{1 - 4 / 3 x (1 - x)}.
\end{align}
For $\xi(s)$, $G(s)$, and $\phi(s)$, the definitions by \cite{StanevLPM, PhysRev.103.1811} are used, which are given by
\begin{align}
  \xi(s) &\approx \xi(s^\prime) =
  \begin{cases}
    2 & \text{if $s^\prime < s_1$}, \\
    1 + h - \frac{0.08 (1 - h) (1 - (1-h)^2)}{\ln{(s_1)}} & \text{if $s_1 \leq s^\prime < 1$}, \\
    1 & \text{if $s^\prime \geq 1$},
  \end{cases}
\end{align}
\begin{align}
  G(s) &=
  \begin{cases}
    3\psi(s) - 2\phi(s) & \text{if $s < \num{0.710390}$}, \\
    36s^2 / \left(36s^2 + 1 \right) & \text{if $\num{0.710390} \leq s < \num{0.904912}$}, \\
    1 - 0.022s^{-4} & \text{if $s \geq \num{0.904912}$},
  \end{cases}
\end{align}
\begin{align}
  \psi(s) &= 1 - \exp{\left\{ -4s - \frac{8s^2}{1 + 3.936s + 4.97s^2 - 0.05s^3 + 7.5 s^4} \right\}},
\end{align}
\begin{align}
  \phi(s) &=
  \begin{cases}
    1 - \exp{\left\{ -6s \left(1 + (3 - \pi) s\right) + \frac{s^3}{ 0.623 + 0.796s + 0.658 s^2} \right\}} & \text{if $s < \num{1.54954}$}, \\
    1 - 0.012 s^{-4} & \text{if $s \geq \num{1.54954}$},
  \end{cases}
\end{align}
with
\begin{align}
  s &= \frac{s^\prime}{\sqrt{\xi(s^\prime)}}, & s^\prime &= \frac{1}{8} \sqrt{\frac{E_\text{LPM}}{E x ( 1 - x)}}, & s_1 &= \frac{\sqrt{2} Z^{\sfrac{2}{3}}}{B^2}, \\ E_\text{LPM} &= \frac{2 \alpha (m_e c^2)^2 X_0}{\pi \hbar c}, & h &= \frac{\ln{(s^\prime)}}{\ln{(s_1)}}, & D_n &= 1.54 A^{0.27}.
\end{align}
Here, $X_0$ is the radiation length, and $B$ is the radiation logarithm constant.
The effect of the LPM effect on pair production can be seen in Figure \ref{fig:photon_dndx}.

\subsection{$\mu^+\mu^-$ Pair Production}
\label{sec:muonpairproduction}
Muon pair production describes the conversion of a photon into a muon-antimuon pair in the field of an atomic nucleus.
While the process is suppressed by a factor of approximately $\left(  m_e / m_\mu \right)^2$ compared to the cross section of electron-positron pair production, it provides a significant contribution to the number of muons in electromagnetic air showers.

\begin{figure}
    \centering
    \includegraphics[width=0.8\textwidth]{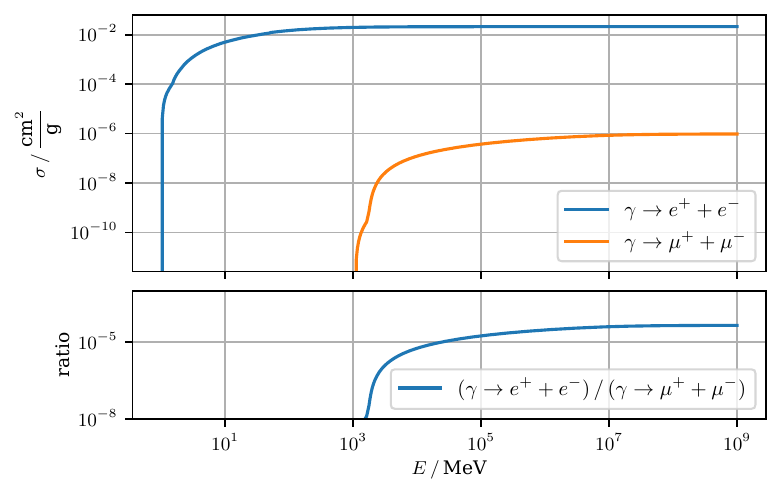}
    \caption{Total cross section of muon pair production in air, compared to the total cross section of electron-positron pair production according to \eqref{eqn:crosssection_pairproduction_kochmotz}.}
    \label{fig:photomupair_cross}
\end{figure}

Using crossing symmetry, the differential cross section for muon pair production is obtained from the differential bremsstrahlung cross section of muons \cite{Kelner:288828} and is given by
\begin{align}
  \frac{\mathrm{d}\sigma}{\mathrm{d}x} &= 4 Z^2 \alpha \left( r_e \frac{m_e}{m_\mu} \right)^2 \Phi(\delta) \left[ 1 - \frac{4}{3} (x - x^2) \right],
\end{align}
with the definitions
\begin{align}
  \Phi(\delta) &= \underbrace{\ln \left( \frac{B Z^{\sfrac{-1}{3}} m_\mu / m_e}{1 + B Z^{\sfrac{-1}{3}} \sqrt{e} \delta / m_e } \right)}_{\Phi_0} - \underbrace{\ln\left( \frac{D_n}{1 + \delta (D_n \sqrt{e} - 2) / m_\mu} \right)}_{\Delta_\text{n}},
\end{align}
where $B$ is the material-dependent logarithm constant \cite{Kelner:1999uj} and
\begin{align}
  x &= \frac{E_{\mu^-}}{E}, & \delta &= \frac{m_\mu^2}{2 E x (1 - x)}, & D_n &= 1.54 A^{0.27}.
\end{align}
The contribution of atomic electrons is taken into account with 
\begin{align}
  \Phi(\delta) &\rightarrow \Phi(\delta) + \frac{1}{Z} \left[ \ln\left( \frac{m_\mu / \delta}{\delta m_\mu / m_e^2 + \sqrt{e}} \right) - \ln\left( 1 + \frac{1}{\delta \sqrt{e} B^\prime Z^{\sfrac{-2}{3}} / m_e} \right) \right],
\end{align}
where $B^\prime$ is the inelastic radiation logarithm \cite{Kelner:1999uj}.
The effect of the inelastic nuclear form factor is taken into account with
\begin{align}
  \Delta_\text{n} &\rightarrow \left( 1 - \frac{1}{Z} \right) \Delta_\text{n}
\end{align}
for $Z>1$.
The kinematic limits in $x$ are defined as 
\begin{align}
  \frac{1}{2} \left( 1 - \sqrt{1 - 2 \sqrt{e} m_\mu / E } \right) \leq x \leq \frac{1}{2} \left( 1 + \sqrt{1 - 2 \sqrt{e} m_\mu / E } \right).
\end{align}
Figure \ref{fig:photomupair_cross} shows the total cross section for muon pair production, compared to electron-positron pair production.

\subsection{Photoelectric Interaction}
\label{sec:photoelectric}
Photoelectric interaction defines the absorption of a photon by an atom with the ejection of a formerly bound electron. 
In general, the cross section depends on the atomic structure of the target atom; however, if the energy of the photon is not near the energy of an absorption edge (of the order of a few keV to a few tens of keV), the cross section can be written in the Born approximation as
(\cite{sauter1931photoeffect1,sauter1931photoeffect2}, cited after \cite{heitler1954})
\begin{multline}
  \label{eqn:photoeffect}
  \sigma_\text{ph.el.} = \sigma_\text{T} \frac{3}{2} Z^5 \alpha^4
    \left(\frac{m_e}{\omega}\right)^5 (\gamma^2 - 1)^{3/2} \\
  \times \left[\frac{4}{3} + \frac{\gamma (\gamma - 2)}{\gamma + 1}
    \left(1 - \frac{1}{2 \gamma \sqrt{\gamma^2 - 1}}
    \ln \frac{\gamma + \sqrt{\gamma^2 - 1}}
    {\gamma - \sqrt{\gamma^2 - 1}}\right)\right],
\end{multline}
where $\omega$ is the incoming photon energy, 
\begin{equation}
  \gamma = 1 + \frac{\omega - I}{m_e}
\end{equation}
is the Lorentz factor of the ejected electron, and $\sigma_\text{T} =
(8 \pi/3) r_e^2$ is the Thomson cross section. $I = Z^2 \alpha^2 m_e/2$
denotes the ionization energy of the $K$-shell electron.

The result above uses the Born approximation, i.\,e., plane wave functions.
Therefore it applies mainly to light elements. The argument of the
logarithm can be rewritten as
\begin{equation}
  \frac{\gamma + \sqrt{\gamma^2 - 1}}{\gamma - \sqrt{\gamma^2 - 1}} =
  \frac{(\gamma + \sqrt{\gamma^2 - 1})^2}
  {\gamma^2 - (\sqrt{\gamma^2 - 1})^2} = (\gamma + \sqrt{\gamma^2 - 1})^2
\end{equation}
to avoid numerical problems at $\gamma \gg 1$.

In \cite{sauter1931photoeffect2}, a correction factor for the nonrelativistic region is
calculated
\begin{equation}
  F = |\Gamma (2 + n)|^2 e^{i \pi n - 4 i n \arctan (1/i n)}
\end{equation}
with $n = \alpha Z/i \beta$. This can be rewritten as
\begin{equation}
  F = \left[1 + \left(\frac{Z \alpha}{\beta}\right)^2 \right]
    \frac{\alpha Z \pi/\beta}{\sinh (\alpha Z \pi/\beta)}
    \exp \left[\frac{Z \alpha}{\beta} \left(\pi - 4 \arctan \frac{\beta}{\alpha Z}
    \right)\right].
\end{equation}
Including this prefactor improves the agreement in comparison to more detailed calculations.

The cross section in \eqref{eqn:photoeffect} describes the dominant absorption on the $K$-shell
electron. An empirical formula from \cite{hubbell1969} for the ratio of the
$K$-shell and the total photoelectric cross section is given by
\begin{equation}
  \frac{\tau_{pe}}{\tau_K} = 1 + \num{0.01481} \ln^2 Z - \num{0.000788}
    \ln^3 Z.
\end{equation}
This parametrization is used as a correction factor to \eqref{eqn:photoeffect} to obtain a better description of the total photoelectric cross section.

\section{Secondary Parametrizations} \label{sec:secondary_params}

\subsection{Bremsstrahlung by electrons and positrons}
\label{sec:secondaries_brems}

The implemented \texttt{KochMotzSampling} method to sample the production angle of bremsstrahlung photons is based on the double differential cross section by Koch and Motz \cite{RevModPhys.31.920}.
The underlying rejection sampling method is adapted from \cite{egs4} and described in more detail in \cite{pirs0203}.
Firstly, a candidate scattering angle $\hat{\theta}$ is sampled via
\begin{equation}
	\label{eqn:theta_hat_brems}
	\hat{\theta} = \frac{m_e}{E} \sqrt{\frac{\xi_1}{1 - \xi_1 + \left(\frac{m_e}{\pi E}\right)^2}},
\end{equation}
with a random number $\xi_1 \in \left[0, 1\right)$.
Next, the normalization of the rejection function is calculated from
\begin{align}
	N_\text{r} &= 1 \,/\, \text{max}\left[ g(0), g(1), g\left(\frac{\pi^2 E^2}{m_e^2}\right) \right]
\end{align}
with the definitions
\begin{align}
	g(x) &= 3\left(1 + r^2\right) - 2r - \left( 4 + \ln(m(x)) \right) \left( \left(1 + r^2\right) - \frac{4 x r}{(1 + x)^2} \right),\\
	m(x) &= \left( \frac{m_e \left(1 - r \right)}{2 r E} \right)^2 + \left( \frac{Z^{\sfrac{1}{3}}}{111 (1 + x)}  \right)^2,
\end{align}
and $r = 1 - v$.
The scattering angle $\hat{\theta}$ is accepted when the condition
\begin{align}
	\xi_2 \leq N_\text{r} \cdot g\left( \frac{E^2 \cdot \hat{\theta}^2}{m_e^2} \right)
\end{align}
is fulfilled, using an additional random number $\xi_2$.
If this is not the case, the process is repeated from equation \eqref{eqn:theta_hat_brems}, using a new set of random numbers $(\xi_1, \xi_2)$.

\subsection{Electron-positron pair production}
\label{sec:secondaries_epair}

An approximative double differential cross section describing the angular distribution of electron-positron pair production based on \cite{tsai1974} is given by
\begin{equation}
  \label{eqn:angular_distribution}
  \begin{split}
  \frac{\mathrm{d}^2\sigma}{\mathrm{d}{\theta}\mathrm{d}{p}} &= \frac{2 \alpha^3 E_{-}^2}{\pi E_{\gamma} m_e^4} \sin(\theta) \biggl[ \left( \frac{2 x (1-x)}{(1 + l)^2} - \frac{12 l x (1-x)}{(1 + l)^4} \right) G_2(\infty) \\ &\quad+ \left( \frac{2 x^2 - 2x + 1}{(1 + l)^2} + \frac{4 l x (1-x)}{(1 + l)^4} \right) \left( X - 2 Z^2 f(z) \right) \biggr],
  \end{split}
\end{equation}
with the angle $\theta$ between the initial photon of energy $E_\gamma$ and produced electron with energy $E_-$, and
\begin{align*}
  x &= \frac{E_{-}}{E_{\gamma}}, & l &= \frac{E_-^2 \theta^2}{m_e^2}, & G_2(\infty) = Z^2 + Z.
\end{align*}
The function $X$, describing the influence of atomic form factors, varies with $Z$ and is defined in \cite{tsai1974}.
The angle $\theta$ can be sampled from this differential cross section by solving the integral equation
\begin{equation}
	\label{eqn:angular_tsai}
    \left( \frac{\mathrm{d}\sigma}{\mathrm{d}p} \right)^{-1} \int_{0}^{\theta} \frac{\mathrm{d}^2\sigma}{\mathrm{d}\theta \mathrm{d}p} \, \mathrm{d}\theta = \xi,
\end{equation}
with a random number $\xi \in [0, 1)$.
This method is called \texttt{Tsai}.

An alternative approach is given based on the leading term of the differential cross section in \cite{RevModPhys.41.581}, which is given by
\begin{align}
	\frac{\mathrm{d}\sigma}{\mathrm{d}\theta_\pm} \propto \frac{\sin{(\theta_\pm)}}{2 p_\pm \left(E_\pm - p_\pm \cos{(\theta_\pm)}\right)^2}.
\end{align}
The deflection angle $\theta$ can be sampled with \cite{Hirayama:2005zm}
\begin{align}
	\label{eqn:sauter_sampling}
	\cos{(\theta_\pm)} = \frac{E_\pm (2 \xi - 1) + p_\pm}{p_\pm (2 \xi - 1) + E_\pm},
\end{align}
using a random number $\xi \in [0, 1)$.
This method is called \texttt{Sauter}.
\section{Default parametrizations}
\label{sec:default_cross}

When the method \texttt{GetStdCrossSections()} (\texttt{C++}) or \texttt{make\_std\_crosssection()} (Python) is called, 
PROPOSAL returns a list of cross sections, using parametrizations that are reasonable for the given particle type.
The specific parametrizations are listed in tables \ref{tab:defaults_electron}, \ref{tab:defaults_positron}, 
\ref{tab:defaults_photon}, and \ref{tab:defaults_muon_tau}. 

Furthermore, the parametrizations for the stochastic deflections are presented in Table~\ref{tab:deflection_params_muon}.
Note that for stochastic deflections, currently, only parametrizations valid and tested for muons are provided.

\begin{table}
	\footnotesize
	\centering
	\caption{Default energy loss parametrizations for electrons.}
	\label{tab:defaults_electron}
	\begin{tabularx}{\textwidth}{l X X}
	\hline
	Interaction type & Parametrization name & Reference \\
	\hline
	Bremsstrahlung & \texttt{ElectronScreening} & \ref{sec:bremsstrahlung} \\
	$e^- e^+$ pair production & \texttt{ForElectronPositron} & \ref{sec:electron_positron_pairproduction} \\
	Ionization & \texttt{BergerSeltzerMoller} & \ref{sec:ionization} \\
	Photonuclear & \texttt{AbramowiczLevinLevyMaor97} & \cite{Abramowicz:1997ms, butkevich_shadowing} \\
	\hline
	\end{tabularx}
\end{table}
\begin{table}
	\footnotesize
	\centering
	\caption{Default energy loss parametrizations for positrons.}
	\label{tab:defaults_positron}
	\begin{tabularx}{\textwidth}{l X X}
	\hline
	Interaction type & Parametrization name & Reference \\
	\hline
	Bremsstrahlung & \texttt{ElectronScreening} & \ref{sec:bremsstrahlung} \\
	$e^- e^+$ pair production & \texttt{ForElectronPositron} & \ref{sec:electron_positron_pairproduction} \\
	Ionization & \texttt{BergerSeltzerBhabha} & \ref{sec:ionization} \\
	Photonuclear & \texttt{AbramowiczLevinLevyMaor97} & \cite{Abramowicz:1997ms, butkevich_shadowing} \\
	Annihilation & \texttt{Heitler} & \ref{sec:annihilation} \\
	\hline
	\end{tabularx}
\end{table}
\begin{table}
	\footnotesize
	\centering
	\caption{Default energy loss parametrizations for photons.}
	\label{tab:defaults_photon}
	\begin{tabularx}{\textwidth}{l X X}
	\hline
	Interaction type & Parametrization name & Reference \\
	\hline
	Pair production & \texttt{KochMotz} & \ref{sec:kochmotz} \\
	Compton & \texttt{KleinNishina} & \ref{sec:compton} \\
	Photoproduction & \texttt{Rhode} & \cite{Rhode1993} \\
	Photoeffect & \texttt{Sauter} & \ref{sec:photoelectric} \\
	\hline
	\end{tabularx}
\end{table}

\begin{table}
	\footnotesize
	\centering
	\caption{Default energy loss parametrizations for muons and taus.}
	\label{tab:defaults_muon_tau}
	\begin{tabularx}{\textwidth}{l X X}
	\hline
	Interaction type & Parametrization name & Reference \\
	\hline
	Bremsstrahlung & \texttt{KelnerKokoulinPetrukhin} & \cite{Kelner:288828} \\
	$e^- e^+$ pair production & \texttt{KelnerKokoulinPetrukhin} & \cite{osti_4563918, Kelner:1998mh} \\
	Ionization & \texttt{BetheBlochRossi} & \cite{bethe, 10.1119/1.1933408} \\
	Photonuclear & \texttt{AbramowiczLevinLevyMaor97} & \cite{Abramowicz:1997ms, butkevich_shadowing} \\

	\hline
	\end{tabularx}
\end{table}

\begin{table}
	\footnotesize
	\centering
	\caption{Default and optional stochastic deflection parametrizations for muons.}
	\begin{tabularx}{\textwidth}{l X X}
	\hline
	Interaction type & Parametrization name & Reference \\
	\hline
	Bremsstrahlung & \texttt{BremsTsaiApproximation} & \ref{sec:defl_brems_geant4} \\
	\quad(optional) & \texttt{BremsGinneken} & \ref{sec:defl_brems_ginneken}\\
	$e^- e^+$ pair production & \texttt{EpairGinneken} & \ref{sec:defl_epair_ginneken} \\
	Ionization & \texttt{IonizNaive} & \ref{sec:defl_ioniz_naive} \\
	Photonuclear & \texttt{PhotoBorogPetrukhin} & \ref{sec:defl_photonuclear_geant4} \\
	\quad(optional) & \texttt{PhotoGinneken} & \ref{sec:defl_photonuclear_ginneken} \\
	\hline
	\end{tabularx}
	\label{tab:deflection_params_muon}
\end{table}

\clearpage
\bibliographystyle{elsarticle-num}
\bibliography{references}







\end{document}